\documentclass[11pt,graphicx,amsmath]{article}
\usepackage{amsmath}
\usepackage{graphicx}
\usepackage{bm}

\usepackage{amssymb}

\title{ Analytic Spectra of CMB Anisotropies and Polarization
Generated by Scalar Perturbations in Synchronous Gauge }

\author{Z. Cai$^{1,2,3}$\thanks{caiz at email.arizona.edu},
               Y. Zhang$^{1}$ \thanks{yzh at ustc.edu.cn}\ ,\\
       $^1$   \small Key Laboratory for Researches in Galaxies and Cosmology, \\
      \small Department of  Astronomy,  University of Science and Technology of China, \\
     \small Hefei, Anhui, 230026,  China\\
         $^2$ \small   Department of Physics,
          University of Arizona, Tucson, AZ 85721, USA \\ 
          $^3$ \small Steward Observatory, University of Arizona, Tucson, AZ, 85721, USA \\   }
\date{}

\begin{document}

\maketitle
\baselineskip=19truept

\newcommand{\B}[1]{\textcolor{blue}{#1}}
\newcommand{\R}[1]{\textcolor{red}{#1}}

\def\gsim{\;\rlap{\lower 2.5pt  \hbox{$\sim$}}\raise 1.5pt\hbox{$>$}\;}
\def\lsim{\;\rlap{\lower 2.5pt  \hbox{$\sim$}}\raise 1.5pt\hbox{$<$}\;}
\def\edth{\;\raise1.0pt\hbox{$'$}\hskip-6pt\partial\;}
\def\baredth{\;\overline{\raise1.0pt\hbox{$'$}\hskip-6pt \partial}\;}

\def\be{\begin{equation}}
\def\ee{\end{equation}}
\def\ba{\begin{eqnarray}}
\def\ea{\end{eqnarray}}
\def\nn{\nonumber}
\def\bt{\bm{\theta}}
\def\cf{\mathcal{F}}
\def\ch{\mathcal{H}}

\sf

\begin{center}
\Large  Abstract
\end{center}

\begin{quote}

 \sf
\baselineskip=19truept

The temperature anisotropies and polarization of
the cosmic microwave background radiation (CMB)
not only {serve as indispensable cosmological probes,}
but also provide a unique channel to  detect
 relic gravitational waves (RGW) at very long wavelengths.
Analytical studies of the anisotropies and polarization
improve our understanding of various cosmic processes
and  help to separate the contribution of RGW from that of density perturbations.

We present a detailed analytical
calculation of CMB temperature anisotropies $\alpha_k $
and polarization $\beta_k$
generated by scalar metric perturbations in synchronous gauge,
parallel to our previous work with RGW as a generating source.
 This is realized primarily by
an analytic time-integration of
 Boltzmann's equation,
yielding the closed forms of $\alpha_k $ and $\beta_k$. 
Approximations, such as the tight-coupling approximation for photons
a prior to the recombination and the long wavelength limit for scalar 
perturbations are used.
The residual gauge modes in scalar perturbations are analyzed 
and a proper joining condition of scalar perturbations
at the radiation-matter equality is chosen,
ensuring the continuity of energy perturbation.

The resulting analytic expressions
of the multipole moments of polarization $a^E_l$,
and of temperature anisotropies  $a^T_l$  
are explicit functions of the scalar perturbations,
 recombination time,
 recombination width,
  photon free streaming damping factor,
 baryon fraction, initial amplitude,
 primordial scalar spectral index, and the running index.
These results show that
a longer recombination width yields higher amplitudes of polarization on large scales
and more damping on small scales,
and that a late recombination time shifts
the peaks of  $C^{XX'}_l$ to larger angular scales.

Calculations show that
$a^E_l$ is generated in the presence of
the quadrupole $\alpha_2$ of temperature anisotropies via scattering,
both having similar structures and being smaller than the total $a^T_l$,
which consists of the contributions from
the monopole, dipole, quadrupole, and Sachs-Wolfe terms as well.
The origin of the two bumps in $C^{EE}_l $ on large angular scales
is found to be due to the time derivative of the monopole of
temperature anisotropies.
Furthermore,
$ a^E_l$ together with $ a^T_l$ demonstrates explicitly that
the peaks of  $C^{EE}_l $ and $C^{TT}_l $ alternate in $l-$ space.
These results substantially extend earlier analytic work.

The analytic spectra $C^{XX'}_l$  agree with the numerical ones
and with those observed by WMAP on large scales ($l \lesssim 500$),
but deviate considerably from the numerical results on smaller scales,
showing the limitations of our approximate analytic calculations.
Several possible improvements are pointed out for further studies.

\end{quote}

\maketitle

\noindent PACS number: 98.70.Vc,  98.80.-k,  98.80.Jk,

\noindent Key words: cosmic microwave background radiation, scalar
              perturbations, polarizations

\newpage
%\twocolumn
\large

\baselineskip=19truept

\begin{center}
{\em\Large 1. Introduction}
\end{center}

By confronting predictions of theoretical cosmological models
with the data on the CMB by the observations,
such as BOOMERANG \cite{Boomerang}, MAXIMA \cite{MAXIMA},  DASI \cite{DASI},
WMAP \cite{WMAP1-over,Peiris,WMAP3,WMAP5,WMAP7,Komatsu},
 Archeops \cite{Archeops},
CBI \cite{CBI}, QUaD \cite{QuaD}, BICEP \cite{BICEP}  etc,
several important cosmological parameters of the standard Big Bang model
have been directly measured or constrained.
These studies have been instrumental for rapid progresses
toward understanding of the evolution of the Universe,
and for the advent of an epoch of precise cosmology.

On the side of theory,
these achievements have been possible
through detailed computations of the spectra
of the CMB temperature anisotropies and polarizations.
Codes of numerical computation,
such as CMBFAST \cite{cmbfast} and CAMB  \cite{Lewis},
give the the spectra $C_l^{XX'}$ of CMB temperature anisotropies and polarizations.
The prominent structure of $C_l^{XX'}$ involves various cosmological parameters,
as it depends upon
several major physical processes
during the cosmic expansion,
such as the inflation, radiation-matter equality,
 recombination, and the reionization as well.
Analytical studies are still indispensable
for understanding how various underlying physical effects give rise to
the observed behavior and
for theoretical interpretations of the observational data.
In particular, the analytical spectra are helpful
in revealing their explicit dependence on the cosmological parameters
and possible degeneracies between them.
So it would be desired to have the
analytical $C_l^{XX'}$ for a better understanding of physics of CMB.

From computational point of view,
the CMB temperature anisotropies and polarizations
are determined by the Boltzamann's equation of the photon gas
in the expanding Universe.
Although a number of ingredients will influence this equation,
two key elements are responsible for the overall features of $C_l^{XX'}$,
i.e.,
Thompson scattering during
the recombination process around a redshift $z\sim1000$
and the metric perturbations $h_{ij}$ of Robertson-Walker spacetime
entering  the equation as the Sachs-Wolfe term \cite{Sachswolfe}.
Generally,
there are two types of metric perturbations as the source:
the scalar (density) perturbations  \cite{Bardeen,Kodama,Mukhanov}
and the tensorial perturbations, i.e., RGW \cite{grishchuk,Ford,zhang}.
Both types can be generated during early stages of the universe, such as
the inflationary expansion.
Among them,
the contribution from scalar perturbations is believed
to be dominant over that from RGW
\cite{basko,Polnarev,Crittenden,
zald,Kosowsky,zaldarriaga,kamionkowski,KeatingTimbie},
characterized by a tensor/scalar ratio $r$ \cite{Liddle,Kosowskyturner}.
For a power law spectrum of the primordial fluctuations,
WMAP5 data alone puts an upper limit on the  ratio
$r<0.55$ (95\% CL) \cite{WMAP5},
while WMAP7 gives
$r <0.49$ (95\% CL) for $\Lambda$CDM+Tensors+Running \cite{WMAP7}.
The recent data of LIGO S5 \cite{LIGO5} with cross-correlation of H1 and L1
gives a constraint $r<0.55$ for the flat primordial tensorial perturbations
with a negligible running index \cite{TongZhang}.
For the case of RGW as the source, Refs. \cite{Pritchard,zhao} derived
the analytical spectrum $C_l^{TT}$ of temperature anisotropies,
and Ref. \cite{xia} gave all four  analytical   spectra $C_l^{EE}$,  $C_l^{BB}$,
  $C_l^{TE}$, as well as $C^{TT}_l$.
Ref. \cite{Xia2} incorporated the reionization process into calculation
and obtained the reionized analytical spectra $C_l^{XX'}$.

Ref.\cite{challinor} presented a fully covariant and gauge-invariant formulation
of Boltzmann's equations.
The analytic calculation of the scalar induced $C^{TT}_l$ was made
in Newtonian (longitudinal) gauge in Refs \cite{HuApJ444, HuNature}.
Ref.\cite{Weinberg} gives an analysis of $C^{TT}_l$ in synchronous gauge,
but the treatment of temperature anisotropies itself
was not enough to separate contributions of monopole, dipole, 
quadrupole, and Sachs-Wolfe terms.
It did not address  the CMB polarizations either.
Ref.\cite{Baskaran} gave a unifying framework in synchronous gauge
to discuss the scalar induced spectra $C^{TT}_l$, $C^{EE}_l$, $C^{TE}_l$
and as well as the RGW induced spectra  $C^{XX'}_l$.
Motivated by possible extractions of signals
of RGW using anti-correlation of $C^{TE}_l$,
attempts were made to estimate qualitatively
the possible forms of multipoles $a_l^T$ of the temperature anisotropies
and $a^E_l$ of the polarization at $l\sim 50$ \cite{Baskaran}.
However, the analysis was still preliminary by lacking of an explicit formula of $a_l^T$,
since the time-integrations of the Boltzmann's equation
as a key procedure was not carried out.
Viewing these, in this paper,
we shall perform a detailed, analytic calculation
of $a_l^T$ and $a_l^E$ induced by the scalar perturbations in synchronous gauge,
and present the analytical spectra $C^{XX'}_l$,
which will be at the same level of accuracy
as the analytical $C_l^{XX'}$ by RGW \cite{zhao,xia,Xia2}.
Aside several new insights into the physics of CMB,
in particular, our resulting cross-correlation spectrum $C^{TE}_l$
has already demonstrated some inaccuracy
in the preliminary analysis of Ref.\cite{Baskaran}.
Therefore, these two sets of analytic spectra together
are more reliable in analyzing and disentangling
the RGW contributions from observational data
\cite{Polnarevmiller,ZhaoBaskaranGrishchuk,LiuLi}.

The synchronous gauge has been often used,
in which the decomposition of generic metric perturbations $h_{ij}$
into the scalar,  vector, and tensorial types is straightforward.
For the scalar metric perturbations,
this gauge is also more efficient in dealing with
the adiabatic and isocurvature initial conditions,
adequate for numerical computations \cite{cmbfast,Lewis}.
In comparison with the conformal Newtonian gauge \cite{HuApJ444},
there are residual gauge freedoms in the synchronous gauge
in the solution of scalar metric perturbations.
By the restricted coordinate transformations
\cite{Press,Ratra,Grishchuk density},
general  solutions become rather involved
for modes of arbitrary wavelengths.
To implement analytical calculations,
we work in the long wavelength approximation.
Besides,
a joining condition of the perturbation modes
at the equality of radiation-matter
will be chosen to ensure
the continuity of the energy perturbations, not of the pressure.

In solving the Boltzmann's equation, One has to
carry out the time-integrations for  $a_l^T$ and $a_l^E$ from the
RD epoch up to the present. The  visibility function for the
recombination process will appear in the integrations, and can be
approximately fitted by the Gaussian type of functions
\cite{zhao,xia,Xia2,Jorgensen}, in order to obtain the analytical
expressions of $a_l^T$ and $a_l^E$.

The organization of this paper is as follows.
In Section 2, we introduce the convention
of  the decomposition of the scalar metric perturbations $h_{ij}$
into two independent modes $h(\tau)$ and $\eta(\tau)$
in the flat Robertson-Walker metric.
In Section 3,
the Boltzmann's equation of the CMB radiation field
in the Basko-Polnarev's framework is formally solved
in terms of two time-integrations
for the temperature anisotropies $\alpha_k(\tau)$
and polarizations $\beta_k(\tau)$,
respectively.
The integrands consist of
some combinations of the metric perturbations,
the monopole $\alpha_0$,
and the dipole $\alpha_1$ of the temperature anisotropies as well.
The fitting formula for the visibility function involved in the integrand
is introduced.
In Section 4,
in the tight-coupling approximation,
both $\alpha_0$ and $\alpha_1$ are solved
in terms  of the metric perturbations.
In Section 5,
the time-integrations are carried out,
yielding the analytical expressions
of  $a_l^T (\tau)$ and $a_l^E (\tau)$, respectively.
In Section 6,
we will remove
the gauge modes from the scalar metric perturbations
for the RD and the MD eras,
make a joining connection of the  perturbations
at the radiation-matter equality,
and choose the proper initial conditions
for the perturbations.
In Section 7,
we present the final analytical spectra
$C_l^{TT} $, $C_l^{TE} $, and $C_l^{EE}$,
and compare them with the numerical and the observed results.
Several interesting properties of CMB anisotropies and polarization
are revealed by the analytic spectra.
Section 8
summarizes the main results and discusses  possible future improvements.
The Appendix provides the formulae that
relate the multipole moments $a_l^T$ and $a_l^E$
to $\alpha_k$ and $\beta_k$, respectively.
The unit with  $\hbar=c=k_B= 1$  will be used.

\begin{center}
{\em\Large 2. Scalar Metric Perturbations in Synchronous Gauge}
\end{center}

For a spatially  flat ($k=0$) Robertson-Walker (RW)
space-time, the metric  is
\be \label{metric}
ds^2=a^2(\tau)[-d\tau^2+(\delta_{ij}+h_{ij})dx^idx^j].
\ee
where $a(\tau)$ is the scale factor as a function of the  comoving time  $\tau $.
The normalization of the scale factor is taken such that
$a(\tau_0)=2/H_0$ at the present time $\tau_0$,
where $H_0$ is the Hubble constant.
In our calculation,
the RD and MD stages of the cosmic expansion are involved,
for which the scale factor can be taken as the following form  \cite{Baskaran}:
\be \label{a1}
 a(\tau)=\frac{4}{H_0\sqrt{1+z_{eq}}}\tau, ~~~~ \tau<\tau_{2};
\ee
\be  \label{a2}
 a(\tau)=\frac{2}{H_0}(\tau+\tau_{2})^2,~~~~ \tau\geq \tau_{2},
\ee
respectively,
where $\tau_{2}=1/(2\sqrt{1+z_{eq}})\simeq 0.0085$
is the radiation-matter equality
for $z_{eq}\simeq 3400$ \cite{WMAP1-over},
and   $\tau_0=1-\tau_{2}$.
In this convention,
the recombination time $\tau_d \simeq 0.0216$,
 corresponding  to a redshift $z\sim 1100$.
As will be seen {in Section 3},
the precise value of $\tau_d$ actually depends upon the
baryon fraction $\Omega_b$.
To keep our analytical calculations simple,
we do not include the current accelerating stage,
which will bring some minor modifications
to the CMB spectra.
The metric perturbations
in the synchronous gauge $h_{ij}$ in Eq.(\ref{metric})
can be generally decomposed as
\be
h_{ij}=\frac{1}{3}h\delta_{ij}+h^\parallel_{ij}+h_{ij}^\bot
          + h_{ij}^\top \, .
\ee
Here $h_{ij}^\bot$ is the transverse ($\partial_i \partial_j h_{ij}^\bot=0$),
vector mode, and is usually neglected as it decays with the cosmic expansion.
$h_{ij}^\top$ is the transverse ($\partial_i h_{ij}^\top=0$),
tensorial mode, i.e., RGW.
Its analytic solution and the analytical spectra $C_l^{XX'}$ induced by $h_{ij}^\top$
have been studied before \cite{zhang,xia,Xia2,miao}.
We consider the remaining part,
which is the scalar metric perturbations,
\be \label{scalar}
h_{ij} =\frac{1}{3}h\delta_{ij}
+h^\parallel_{ij} \, ,
\ee
where $h\equiv h^i_i $ is the trace part,
and $h^\parallel_{ij}$ is the traceless and  longitudinal part,
satisfying
\[
\epsilon_{ijk} \partial_j\, \partial_l  h^\parallel_{lk}=0.
\]
It can be expressed in terms of a scalar function,
\be \label{hparallel}
h^\parallel_{ij} =  \left(\partial_i\partial_j
             -\frac{1}{3}\delta_{ij}\nabla^2 \right)\upsilon.
\ee
Thus the density perturbations $h_{ij}$  are described by two scalar functions,
and can be written as a Fourier integration \cite{Baskaran,ma}

\be\label{Fourier}
h_{ij}( {\bf x},\tau)=\int d^3k   e^{i \,\bf{k}\cdot\bf{x}}
    \left( \sum_{s=1,2} {\mathop p\limits^s}_{ij} ~ {\mathop h\limits^s}_k(\tau)
       \right), \,\,\,\, {\bf k}=k\vec{e}_k,
\ee
where ${\mathop h\limits^1}_k(\tau) $  and ${\mathop h\limits^2}_k(\tau) $
are the two scalar functions introduced,
and
\be \label{polarization}
{\mathop p\limits^1}_{ij}= \vec{e}_{k_i}\vec{e}_{k_j},  ~~~~~
{\mathop p\limits^2}_{ij}=  (\vec{e}_{k_i}\vec{e}_{k_j}-\frac{1}{3}\delta_{ij})
\ee
are the two corresponding polarization tensors
for the density perturbations.
If we write
\be \label{h eta}
{\mathop h\limits^1}_k(\tau)  =h_k(\tau), ~~~~~
{\mathop h\limits^2}_k(\tau)  =6\eta_k(\tau),
\ee
then  $h_k(\tau)$ and $\eta_k(\tau)$ are the scalar modes used in Ref.\cite{ma}.
If we write
\be \label{HHl}
{\mathop h\limits^1}_k(\tau)  =-H_l\, _k(\tau)+3H_k(\tau), ~~~~~
{\mathop h\limits^2}_k(\tau)  =-3 H_k(\tau),
\ee
then  $H_k$ and $H_l\,_k$ are identified as
the scalar modes employed in Refs. \cite{Grishchuk density,Baskaran},
where small letters
$  h_k$ and $ h_l\, _k  $ were used.
The  sets $(h, \eta)$ in Eq.(\ref{h eta})
is related to the set $(H, H_l)$ in Eq.(\ref{HHl}) as the following
\be   \label{relation}
H=-2\eta,~~~~~~~  H_l=-(h+6\eta),
\ee
where
the sub-index $k$ has been omitted  in the following when no confusion arises.

{  An important property of density perturbations
is that,
a  $\bf k$ mode  of $h_{ij}$ in Eq.(\ref{scalar})
is rotationally symmetric about the $\bf k$ axis.
Let the polar axis $z$ be along $\bf \hat k$.
The $\bf k$ mode of the trace part $\frac{1}{3}h\delta_{ij}$
is isotropic in space,
and the longitudinal part $h_{ij}^{||}$ has only the $zz$ component.
This is also reflected by its polarizations
${\mathop p\limits^1}_{ij}$ and ${\mathop p\limits^2}_{ij}$
given in Eq.(8)
that only depend on the vector $\bf \hat k$,
independent of any vector perpendicular to $\bf  k$.
So the $\bf k$ mode of $h_{ij}$ is independent of the azimuthal angle $\phi$.
In contrast,
a  $\bf k$ mode (plane wave) of GW $h_{ij}^\top$ is transverse,
with  two components $h_{xx}=-h_{yy}$ and  $h_{xy}=h_{yx}$.
Therefore,
$h_{ij}^\top$ is not rotationally symmetric about the $\bf k$ axis,
and does depend on the azimuthal angle $\phi$.
Due to this difference,
as we shall see later,
the $\phi$-independent density perturbation does not
induce the magnetic type of polarization of CMB,
whereas the $\phi$-dependent GW does.}

In order to calculate
the evolution of CMB anisotropies and polarization,
one needs the dynamic evolution of scalar perturbations
$h(\tau)$ and $\eta(\tau)$
that enters the Boltzmann's equation of photons.
However, in the synchronous gauge,
{the solution of $h(\tau)$ and $\eta(\tau)$ contain the residual gauge modes
for both the RD and MD stages, which have to be dealt with later (in Section 6).}

\begin{center}
{\em\Large 3. Boltzmann's equation in   RW spacetime}
\end{center}

The temperature field of CMB is not exactly isotropic,
instead it has anisotropies, which are related to
the metric perturbations  $h_{ij}$ via the Sachs-Wolf term.
Moreover,
the quadrupole component of the temperature anisotropies
will %further induce
{be further induced by} the linear polarizations
via the Thomson scattering during the recombination.
So  the radiation field is
described by the following column vector
\cite{Baskaran,basko,Polnarev,chandrasekar,keating}
\be   \label{n}
\vec{n} =\frac{1}{2\nu^3} \left( \begin{array}{l}
 I+Q  \\
 I-Q  \\
 -2U \\
 \end{array} \right),
\ee
where $I$ is the intensity of radiation,
and $Q$ and $U$ together describe the
linear polarizations.
The column $\vec{n}$ can be split into two parts
\be
\vec{n}={n}^{(0)} \vec{u}+ {\vec{n}}^{(1)}
\,\,\,\,\,\, {\rm with} \,\,\,\,\,\,
\vec{u}       \equiv
\left( \begin{array}{l}
 1 \\    1 \\     0 \\
 \end{array} \right),
\ee
where ${n}^{(0)}$ is the homogeneous,
isotropic and unpolarized Planck spectrum
in the expanding universe with frequency rescaled by the scale factor
$\widetilde{\nu}=\nu a(\tau)$,
and $\vec{n}^{(1)}=\vec{n}^{(1)}(\tau,x^i,\nu,e^i)$
represents the temperature anisotropies and polarizations
caused by the metric perturbation $h_{ij}$,
and   is a
function of the conformal time $\tau$, the comoving spatial
coordinates $x^i$, the photon frequency $\nu$, and the photon
propagation direction
$e^i=(\sin\theta\cos\phi,\sin\theta\sin\phi,\cos\theta)$.

Parallel to the Fourier expansion of $h_{ij}$ in Eq.(\ref{Fourier}),
$\vec{n}^{(1)}$
is also expanded  into:
\[
{\vec{n}}^{(1)}(\tau, x^i, \nu, e^i)
   =\int d^3k   e^{i \,\bf{k}\cdot\bf{x}}\,
    {\vec{n}}^{(1)}_k (\tau,\nu, e^i).
\]
For each Fourier component $ {\vec{n}}^{(1)}_k (\tau,\nu, e^i)$,
up to the first order of perturbations,
the Boltzmann's equation  is written as
\cite{Baskaran,keating}:
\begin{eqnarray} \label{Boltzmann2}
\left(\frac{\partial }{{\partial \tau }} + q(\tau )
   + ie^i k_i \right) {\vec{n}}^{(1)}_k  (\tau ,\nu,e^i )
  = && \frac{{f(\tilde{ \nu}) n^{(0)} (\tilde{\nu})}}{2}
  \left( e^i e^j \sum_{s=1,2} {\mathop p\limits^s}_{ij}
  \frac{d{\mathop h\limits^s}_k (\tau )} {d\tau }
  -q(\tau )e^i v_i \right)  \vec{u}\nonumber+ \\
 && + \frac{{q(\tau )}}{{4\pi }}\int {d\Omega '}\hat{\bf{P}}
     (e^i ;e'^{j} )\vec{n}^{(1)}_k (\tau , \nu,e'^{j} ),
\end{eqnarray}
where the differential optical depth
$q(\tau)=\sigma_TN_e(\tau) a(\tau)$ with $\sigma_T$ being the Thomson cross
section, and
$N_e(\tau)$ being the comoving number density of free electrons,
$f(\tilde{\nu})=-\frac{\partial \ln n^{(0)}}{\partial \ln \tilde{\nu}}\,$,
$e^i e^j \sum_{s=1,2} {\mathop p\limits^s}_{ij}\,
 and \frac{d}{d\tau }\, {\mathop h\limits^s}_k (\tau )$
being the Sachs-Wolfe term  \cite{Sachswolfe}
reflecting the frequency variation due to $h_{ij}$,
and $ v_i$ is the velocity of scattering
electrons with respect to the chosen synchronous coordinate system.
In the frame associated with the density waves with a wavevector
$\vec{k}/k=(0,0,1)$ in $\hat z$ direction,
one has $e^iv_i=-i\mu v_b$ with   $\mu= \cos \theta$
and $v_b$ being the baryon (electron) velocity,
$e^ie^j{\mathop p\limits^1}_{ij}=\mu^2$,
and $e^ie^j{\mathop p\limits^2}_{ij}= (\mu^2 - 1/3)$,
independent of the azimuthal angle $\phi$.
{
Thus, as is expected,
the Sachs-Wolfe term is $\phi$-independent,
because the $\bf k$ mode of
density perturbation $h_{ij}$ is $\phi$-independent,
as mentioned in Section 2.
Then, the only term in Eq.(\ref{Boltzmann2})
that might possibly depend on $\phi$ is
the scattering term
$\int {d\Omega '} \hat{\bf{P}} (e^i ;e'^{j} )\vec{n}^{(1)}_k (\tau, \nu,e'^{j} )$,
where the $\phi$-dependent part of
the Chandrasekhar matrix $ \hat{\bf{P}} (e^i ;e'^{j} )$
is only through $\sin(\phi'-\phi)$ and $\cos(\phi'-\phi)$ \cite{chandrasekar}.
When one takes $\vec n^{(1)}_k$ to be independent of $\phi$,
the $\phi$-dependent part of $\int_0^{2\pi} d\phi'\hat{\bf{P}} (e^i ;e'^{j} ) $
is vanishing due to
$\int_0^{2\pi} d\phi' \sin(\phi'-\phi)=\int_0^{2\pi} d\phi' \cos(\phi'-\phi)=0$.
Therefore, in the case of density perturbations,
it is consistent to take
$\vec n^{(1)}_k$ independence of $\phi$. This $\phi$-independent property is 
the reason that the magnetic type polarization is not sourced by density perturbations (see Appendix). 
In contrast, for the case of GW,
the term $e^i e^j \sum_{t=1,2} {\mathop p\limits^t}_{ij}
\frac{d{\mathop h\limits^t}_k (\tau )} {d\tau }\propto \cos 2\phi$,
 depending on $\phi$. This $\phi$-dependent property is also responsible 
 for the magnetic type polarization generated by GW (see Appendix).  }

To further decompose Eq.(\ref{Boltzmann2}),
one can follow the treatment of
Basko and Polnarev \cite{basko,Polnarev,keating}
and writes ${\vec{n}}^{(1)}_k$
in the following form:
\be\label{n1}
{\vec{n}}^{(1)}_k(\tau, \nu,   \mu)
      =\frac{f(\nu)n^{(0)}(\nu)}{2}
      \left[  \alpha_k(\tau, \mu)
      \left(
\begin{array}{l}
 1 \\
 1 \\
 0 \\
 \end{array} \right)+\beta_k(\tau,\mu)\left( \begin{array}{l}
 1 \\
 -1 \\
 0 \\
 \end{array} \right)  \right],
\ee
where $\alpha_k$ is the temperature anisotropies and
$\beta_k$ is the polarization.
By comparing Eq.(\ref{n}) and Eq.(\ref{n1}),
it is seen that, for each wavenumber $k$,
$\alpha_k$ is proportional to the anisotropic part of the intensity $I$,
\be \label{Ialpha}
I(\tau, \mu)=\gamma \alpha_k(\tau, \mu)
\ee
with the factor $\gamma\equiv \nu^3 f(\nu) n^{(0)}(\nu) $,
and
$\beta_k(\mu)$ is related  to  the linear polarization $Q$ itself
\be \label{Qbeta}
Q(\tau, \mu)=\gamma \beta_k(\tau, \mu).
\ee
{
Note that,
by Thomson scattering of the unpolarized light
at the last scattering,
 the Stokes parameter $U=0$
for the $\bf k$ mode of density perturbation.
If the metric perturbation is GW,
the form of  ${\vec{n}}^{(1)}_k$ will be more complicated than
Eq.(\ref{n1}),
with all three Stokes parameters
$I=I(\theta,\phi)$, $ Q=Q(\theta,\phi)$, and $  U=U(\theta,\phi)\neq 0$,
depending on both angles $(\theta,\phi)$ \cite{basko,Polnarev,zhao}.
See the Appendix for the details.}

Then Eq.(\ref{Boltzmann2}) is converted into a set of
two coupled first order differential equations
for $\alpha_k$ and $\beta_k$  \cite{Baskaran,keating},
\be  \label{alpha}
\left(  \frac{\partial}{\partial \tau}  +q(\tau)+i k\mu\right)\alpha_k
= \frac{1}{2} \left(\frac{d H}{d\tau}-\mu^2\frac{dH_l}{d\tau}\right)
        +q(\tau)\left(\mathcal {I}_1(\tau)
             +i\mu v_b-\frac{1}{2}P_2(\mu)\mathcal{I}_2(\tau)\right), \\
\ee
\be \label{beta}
 \left(\frac{\partial}{\partial \tau}+q(\tau)+i k\mu\right)\beta_k
    =\frac{1}{2}q(\tau)(1-P_2(\mu))\mathcal{I}_2(\tau),
\ee
where $P_2(\mu)$ is the second order Legendre function, and
\be \label{I_1}
\mathcal{I}_1(\tau)\equiv \frac{1}{2}\int_{-1}^{1}d\mu\alpha_k(\tau,\mu),
\ee
\be \label{I_2}
\mathcal{I}_2(\tau)
        \equiv \frac{1}{2}\int_{-1}^{1}d\mu
        \left[(1-P_2(\mu))\beta_k(\tau,\mu)
             -P_2(\mu)\alpha_k(\tau,\mu)  \right],
\ee
play a role of sources for $\alpha_k$ and $\beta_k$.
On the right hand side of Eq.(\ref{alpha}),
$\frac{1}{2}(\frac{d H}{d\tau}-\mu^2\frac{dH_l}{d\tau})$
is the Sachs-Wolfe term,
which has an counterpart in the case of RGW \cite{zhao,xia, Xia2}.
{In contrast to  the case of RGW,
the second term on the right hand side of Eq.(\ref{alpha}) is
a new collision term,
containing $v_b$,  $\mathcal{I}_1$, and $\mathcal{I}_2$,
which all contribute to the temperature anisotropies $\alpha_k$.}
From Eq.(\ref{beta}) one sees that $\mathcal{I}_2$
enters the collision term and plays the role of
the source for the polarization $\beta_k$.
To  solve the set of equations (\ref{alpha}) and (\ref{beta}),
one needs the quantities of $q(\tau)$, $H(\tau)$, $H_l(\tau)$,
$v_b(\tau)$, $\mathcal{I}_1$ and $\mathcal{I}_2$,
which will be determined {in Section 5 and Section 6}.

Although formally similar
to the case of RGW \cite{zhao,xia, Xia2},
 Eqs.(\ref{alpha}) and (\ref{beta})   are more complicated.
There are residual gauge modes contained
in the solutions of $H(\tau)$ and $H_l(\tau)$
for the RD and MD eras
that have to be removed
before one can actually calculate $\alpha_k$ and $\beta_k$.
Also,  the collision term $q(\tau)( ... )$
on the right hand side of Eq.(\ref{alpha})
is absent in the case of RGW, and
needs some extra, proper treatments here.

\begin{figure}
 \resizebox{120mm}{!}{\includegraphics{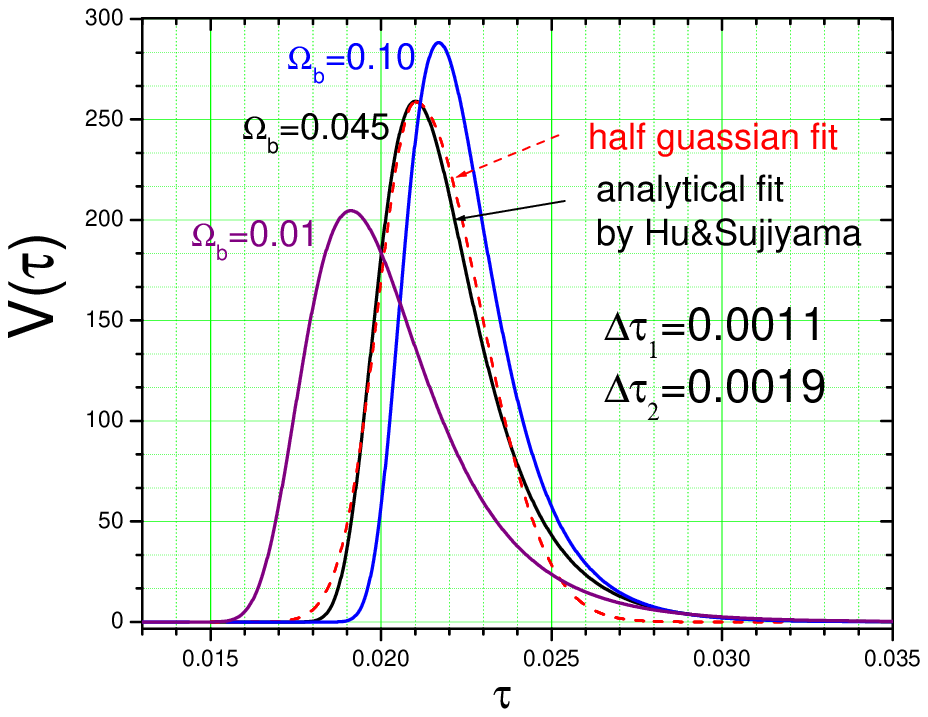}  }
\caption{  \label{fig1} \small
The visibility function $V(\tau)$ for the decoupling.
The solid lines are given by
the analytic formulae for different $\Omega_b$
from Ref.\cite{HuApJ444}.
The dash line is the fitting by
two pieces of half Gaussian functions as in Eq.(\ref{halfgaussian1}).
   }
 \end{figure}

We proceed to write down
the formal solutions to Eqs.(\ref{alpha}) and (\ref{beta})
 as the following time integrations
\be \label{alpha_n}
\alpha_k(\tau,\mu)
    =\int_{0}^{\tau}d\tau'e^{-\kappa(\tau,\tau')-i \mu\,  k(\tau-\tau')}
 \left[ \frac{1}{2}(\frac{dH}{d\tau'}-\mu^2\frac{dH_l}{d\tau'})
+q(\tau')(\mathcal{I}_1(\tau')+i\mu v_b
        -\frac{1}{2}P_2(\mu)\mathcal{I}_2(\tau'))\right],
\ee
\be \label{beta_n}
\beta_k(\tau,\mu)=\frac{1}{2}(1-P_2(\mu))
    \int_{0}^{\tau}d\tau'q(\tau')
     e^{-\kappa(\tau,\tau')-i \mu\, k (\tau-\tau')}\mathcal{I}_2(\tau'),
\ee
respectively,
where
\be \label{kappa'}
\kappa(\tau,\tau') \equiv\int_{\tau'}^{\tau}d\tau''q(\tau'')
=\kappa(\tau')-\kappa(\tau),
\ee
and the optical depth for the recombination
\be \label{kappa}
\kappa(\tau)\equiv
\kappa(\tau_0,\tau)=\int_{\tau}^{\tau_0}d\tau' q(\tau')
\ee
from the present time $\tau_0$ back to an earlier time $\tau$,
whose time derivative yield  the differential optical depth
\be \label{q}
q(\tau)= - \dot{\kappa}(\tau) \equiv - \frac{d \kappa(\tau)}{d\tau}.
\ee
From $\kappa(\tau)$ and $q(\tau)$ follow the visibility function
  \cite{HuApJ444,peebles,peebles2,Sunyaev,jones-wise,Peebles-Yu,Mortonson}
\be \label{V}
    V(\tau) =\frac{d}{d\tau} e^{-\kappa(\tau)}=q(\tau)e^{-\kappa(\tau)}.
\ee
and the exponential function
\be   \label{ekappa}
e^{-\kappa(\tau)}=\int^\tau_{\tau_0}V(\tau)d\tau.
\ee
The quantities $q(\tau)$, $\kappa(\tau)$, $e^{-\kappa(\tau)}$,
and  $V(\tau)$ are equivalent
in describing the recombination process.
In principle,
for a given cosmological model with a known $a(\tau)$,
once the number density of free electrons $N_e(\tau)$ given explicitly
for the detailed recombination process,
one can calculate directly
the differential optical depth $q(\tau)$,
$\kappa(\tau)$, and $V(\tau)$
from their definitions
\cite{xia,Xia2,peebles2}.
The visibility function $V(\tau)$
has a statistical interpretation as the probability that
a CMB photon we observe was last scattered at an earlier time $\tau$,
so that it satisfies the normalization condition
\be \label{vnormalization}
\int_0^{\tau_0} V(\tau)d\tau=1.
\ee
Here we do not consider the reionization process \cite{Xia2},
which would bring another term into
the integrand in Eq.(\ref{vnormalization}).
The recombination process  and the corresponding $V(\tau)$
depend on the matter fraction $\Omega_m$
and the baryon fraction $\Omega_b$.
As depicted in  Fig.\ref{fig1},
$V(\tau)$ is rather sharply distributed around the recombination time $\tau_d$,
and, among other things,
 its dependence  upon the baryon fraction $\Omega_b$
 is such that a greater $\Omega_b$
yields a slightly larger recombination time $\tau_d$.
In our context,
by $\kappa(\tau_d)=1$.
According to \cite{HuApJ444}, one has
\be
\tau_d+\tau_2 \simeq 10^{-3/2} \Omega_b^{0.215/(16+1.8\ln \Omega_b)}.
\ee
In practice, $V(\tau)$ is often be approximated
by some fitting formulae \cite{jones-wise,HuApJ444,xia,Xia2}.
For the purpose of analytic calculations of CMB polarization,
$V(\tau)$ was further simplified by
 a single Gaussian type of function \cite{zald,Pritchard}
\be  \label{v}
V(\tau)     =V(\tau_d) \exp\left(-\frac{(\tau-\tau_d)^2}{2
                 \Delta\tau_d^2}\right),
\ee
where $V(\tau_d) = (\sqrt{2\pi}\Delta\tau_d)^{-1}$
required by the normalization of Eq.(\ref{V}),
and $\Delta\tau_d$ is the half width and reflects
the thickness of recombination.
$\Delta\tau_d$ also depends on
the baryon fraction  $\Omega_b$,
and a larger  $\Omega_b$ yields a slightly narrower $\Delta\tau_d$.
It can be approximately fitted by
\be \label{Deltatau}
\Delta\tau_d \simeq  10^{-4}(8-2.33\ln\Omega_b).
\ee
For the redshift thickness of the recombination
$\Delta z\simeq 195\pm 2$ by WMAP1 \cite{WMAP1-over},
the corresponding conforming time width
is $2 \Delta\tau_d \simeq 0.003 $ for $\Omega_b=0.045$.
As adopted in the case of RGW \cite{zhao,xia,Xia2},
for a better approximation,
$V(\tau)$ has also been  fitted by two pieces of half Gaussian functions
\be\label{halfgaussian1}
V(\tau)=   \left\{
        \begin{array}{ll}
V(\tau_d) \exp\left(-\frac{(\tau-\tau_d)^2}{2
\Delta\tau_{d1}^2}\right),        ~~~(\tau\leq\tau_d) ,   \\
V(\tau_d) \exp\left(-\frac{(\tau-\tau_d)^2}{2
\Delta\tau_{d2}^2}\right),         ~~~(\tau>\tau_d),
          \end{array}
     \right.
\ee
where $\Delta\tau_{d1}=0.0011$ and $\Delta\tau_{d2}=0.0019$
for $\Omega_b=0.045$
and $(\Delta\tau_{d1}  + \Delta\tau_{d2})/2= \Delta\tau_{d}$.
It has been  checked that the errors between
Eq.(\ref{halfgaussian1}) and the approximate formulae proposed in
Refs. \cite{HuApJ444,jones-wise} are very small, $< 4\%$ for $\tau>\tau_d$.
Eq.(\ref{halfgaussian1})
improves the description of
visibility function by $\sim 10\%$ in accuracy
over Eq.(\ref{v}),
and at the same time allows
an analytical calculation of $C_l^{XX'}$.

Now back to $\alpha_k(\tau,\mu)$ and $\beta_k(\tau,\mu)$
in Eqs.(\ref{alpha_n}) and (\ref{beta_n}).
To get rid of their dependence of $\mu = \cos\theta$,
one proceeds to expand them in terms of the Legendre functions:
\be \label{alpha expansion}
\alpha_k(\tau,\mu)=\sum_l(-i)^l\alpha_l(\tau)P_l(\mu),
\ee
\be \label{beta expansion}
\beta_k(\tau,\mu)=\sum_l(-i)^l\beta_l(\tau)P_l(\mu),
\ee
with the  multipole moments given by
\be \label{alpha_l}
\alpha_l(\tau)
        =i^l\frac{2l+1}{2}\int_{-1}^{+1}d\mu\alpha_k(\tau,\mu)P_l(\mu),
\ee
\be \label{beta_l}
\beta_l(\tau)=i^l\frac{2l+1}{2}\int_{-1}^{+1}d\mu\beta_k(\tau,\mu)P_l(\mu),
\ee
where the following normalization condition has been used
\be \label{legend1}
\int_{-1}^{1}dxP_l(x)P_{l\,'}(x)= \frac{2}{2l+1}\, \delta_{l l\, '}  \, .
\ee
Inserting Eqs.(\ref{alpha expansion}) and (\ref{beta expansion}) into
Eqs.(\ref{I_1}) and (\ref{I_2}),
carrying  out the angular integration $\int d\mu$ there,
and using the relation $(2l+1)xP_l(x)=(l+1)P_{l+1}(x)+lP_{l-1}(x)$,
the sources  $\mathcal{I}_1$ and $\mathcal{I}_2$ can be expressed in terms of
the multipoles as the following:
\be \label{I1}
\mathcal{I}_1(\tau)=\alpha_0(\tau),
\ee
\be \label{I2}
\mathcal{I}_2(\tau)=\beta_0(\tau)+\frac{1}{5}\beta_2(\tau)
         +\frac{1}{5}\alpha_2(\tau).
\ee
In particular,  Eq.(\ref{I2}) tells
that the quadrupole $\alpha_2$ of the temperature anisotropies
enters $\mathcal I_2$ as a source for the polarization mode $\beta_l(\tau)$.
In the following
we will express $\mathcal{I}_1(\tau)$
and $\mathcal{I}_2(\tau)$
in terms of  the scalar perturbations $h(\tau)$ and $\eta(\tau)$.

\begin{center}
{\em\Large 4. Determination of Integrands
                      for $\alpha_k(\tau)$ and $\beta_k(\tau)$}
\end{center}

The Boltzmann's equation (\ref{alpha}) and
(\ref{beta}) can be written as hierarchical sets of equations
for  the multipole moments $\alpha_l$ and $\beta_l$ as the
following.

For each $l$, multiplying both sides of Eq.(\ref{alpha}) by
$P_l(\mu)$ and integrating over $\frac{1}{2}\int_{-1}^{1}d\mu $,
one arrives at the hierarchical set of equations for $\alpha_l$:
\be \label{alpha_0}
\dot{\alpha}_0=-k\frac{1}{3}\alpha_1
   +\frac{1}{3}\dot{h},
\ee
\be \label{alpha_11}
\dot{\alpha}_1= k(\alpha_0-\frac{2}{5}\alpha_2)-q(\alpha_1 -v_b),
\ee
\be \label{alpha_2}
\dot{\alpha_2}=k(\frac{2}{3}\alpha_1-\frac{3}{7}\alpha_3)
   -\frac{2}{3}(\dot h+6\dot\eta)-q(\tau)(\frac{9}{10}\alpha_2
    -\frac{1}{2}\beta_0-\frac{1}{10}\beta_2),
\ee
\be
\dot{\alpha_l}
   =k(\frac{l}{2l-1}\alpha_{l-1}
     -\frac{l+1}{2l+3}\alpha_{l+1})-q\alpha_l,
     \,\,\,\,\, l\ge 3.
\ee
Note that the monopole $\alpha_0$ {$=\delta T/T$}=$\delta_\gamma/4$,
where $\delta_\gamma=\delta \rho_\gamma/\rho_\gamma$
of the  photon gas,
and the dipole moment $\alpha_1$
represents the velocity of photon gas.
Eq.(\ref{alpha_0}) shows that the metric perturbation $h(\tau)$
induces the generation of anisotropies $\alpha_0$.
Similar treatments of Eq.(\ref{beta})
yield:
\be  \label{beta0}
\dot{\beta_0}=-k\frac{1}{3}\beta_1
   +q(-\frac{1}{2}\beta_0+\frac{1}{10}(\alpha_2+\beta_2)),
\ee
\be
\dot{\beta_1}=k(\beta_0-\frac{2}{5}\beta_2)-q\beta_1,
\ee
\be \label{beta2}
\dot{\beta_2}=k(\frac{2}{3}\beta_1
   -\frac{3}{7}\beta_3)+q(\tau)(\frac{1}{2}\beta_0
   +\frac{1}{10}\alpha_2-\frac{9}{10}\beta_2),
\ee
\be \dot{\beta_l}=k(\frac{l}{2l-1}\beta_{l-1}
           -\frac{l+1}{2l+3}\beta_{l+1})-q\beta_l,
           \,\,\,\,\, l\ge 3.
\ee
As Eq.(\ref{beta0}) demonstrates,
the quadrupole of temperature anisotropies $\alpha_2$ is the major source
for the leading order polarization $\beta_0$ via scattering.
The above hierarchical sets, for both $\alpha$ and $\beta$,
have infinite number of differential equations,
and should be made closed in order to find their solutions.
One takes the cutoff
\be  \label{ansatz}
\alpha_l=0, \,\,\,\,\, \beta_l=0,  \,\,\,  (l \ge 3),
\ee
which is justified for the long wave modes with $k\tau \ll  1$.
Dropping  the small quadrupole $\alpha_2$,
 Eq. (\ref{alpha_11}) reduces to
\be  \label{dot alpha_1}
\dot{\alpha}_1=k\alpha_0 - q(\alpha_1 -v_b).
\ee
The term $-q(\alpha_1 -v_b)$
represents the momentum transfer from the
baryon (electron) into the photon component.
The quantity $1/q$ has the meaning of the mean free path of photons,
and is a small parameter before the recombination.
Before the recombination,
photons and baryons are tightly coupled,
and in the tight-coupling limit $1/q \rightarrow 0$,
Eq. (\ref{dot alpha_1}) implies $\alpha_1 =v_b$
so that photons and baryons behave
like a coupled single fluid \cite{Peebles-Yu}.
But, for a more accurate account for the difference between
photons and baryons,
one keeps up to the order of $1/q$.
To deal with $v_b$,
one needs to use the momentum conservation in Thomson scattering,
i.e., the Euler equation for the electron velocity
( See Eq.(66)  in Ref. \cite{ma})
\be \label{vb}
\dot v_b= -\frac{\dot a}{a} v_b
          +c_s^2 k^2 \delta_b + \frac{q}{R}(\alpha_1- v_b),
\ee
where $R\equiv  3\rho_b/4\rho_\gamma$ is $3/4$ times
the baryon-photon ratio and, for a model $\Omega_b \sim 0.045$,
can still be treated as $ \ll 1$ during the recombination,
and $c_s=1/\sqrt{3(1+R)}$ is the sound speed of the photon gas.
In the tight-coupling limit,
the particle collision rate via Thomson scattering
is much greater than the expansion rate, i.e., $q\gg \dot a/a $.
So, in the long wavelength limit,
Eq.(\ref{vb}) reduces to
\be \label{vb2}
\dot v_b
\simeq \frac{q}{R}(\alpha_1- v_b).
\ee
Now by combination of Eqs.(\ref{alpha_0}), (\ref{dot alpha_1}), and (\ref{vb2}),
one obtains the following
second order differential equation of
the monopole:
\be \label{ddotalpha0}
\ddot{\alpha}_0  +c^2_s k^2\alpha_0   =S(\tau)
\ee
with the source
\be \label{S}
S(\tau)\equiv \frac{1}{3}\ddot{h}=\ddot{H}-\frac{1}{3}\ddot H_l,
\ee
where the second equal sign follows by Eqs.(\ref{relation}).
Note that  $c_s$ appearing in Eq.(\ref{ddotalpha0})
is a function of time
through the ratio $R$.
At the level of our analytical calculation,
$R$ will be  approximately treated as a constant,
and its value will be taken at  $z\sim 1100$
around the recombination.
The general solution of Eq.(\ref{ddotalpha0}) is given by
the following simple form
\be \label{alpha0}
\alpha_0(\tau)=B_1\cos (p\tau) +B_2\sin(p\tau)
   +\int_{0}^{\tau}
\frac{S(\tau')\sin(p\tau-p\tau')}{p}d\tau',
\ee
where $p\equiv c_sk$,
and $B_1$ and $B_2$ are constants, to be fixed by  initial conditions.
As we shall see later in Section 6,
both $B_1$ and $B_2$ can be set to vanish.
Once the scalar perturbation mode $h(\tau)$ is specified,
the monopole  $\alpha_0(\tau)$ follows from Eq.(\ref{alpha0}),
and so does the source $\mathcal{I}_1(\tau)$ via Eq.(\ref{I1}).
We remark that
if the expansion term $-\frac{\dot a}{a} v_b$ in Eq.(\ref{vb}) was kept,
Eq.(\ref{ddotalpha0}) would be modified as the following
\be \label{dotR}
\ddot{\alpha}_0 + \frac{\dot R}{1+R}\dot \alpha_0
    +c^2_s k^2\alpha_0
    = \frac{\dot R}{1+R}\frac{1}{3}\dot h +  S(\tau).
\ee
With $R$ being a time-dependent function,
this differential equation could also be solved,
whose solution would
 differ only slightly from  Eq.(\ref{alpha0})
on large scales under consideration ($l\lesssim 400$).
%To keep our presentation simple,
{In order to get full-analytical formulae, }
we use Eq.(\ref{alpha0}).

Once  $\alpha_0(\tau)$ and $h(\tau)$ are known,
the dipole $\alpha_1$ follows immediately
from Eq.(\ref{alpha_0}),
\be \label{alpha1}
\alpha_1(\tau) =\frac{3}{k}(-\dot{\alpha}_0
          +\frac{1}{3}\dot{h}).
\ee
Let us calculate the source  $\mathcal{I}_2$ for polarization
in Eq.(\ref{I2}).
It is interesting that
a linear combination of Eqs.(\ref{alpha_2}), (\ref{beta0}), and (\ref{beta2}),
with higher order terms {($l \ge$ 3)} in perturbations being  dropped,
leads to the following differential equation
\be
\label{huaxieI2}
\dot{\mathcal{I}}_2 +\frac{3q}{10}\mathcal{I}_2 =M(\tau)
\ee
with
\be \label{M}
M(\tau)\equiv
 -\frac{2}{5} \dot \alpha_0(\tau)
             - \frac{4}{5} \dot{\eta}(\tau)
=-\frac{2}{5}\dot \alpha_0 +\frac{2}{5}\dot H.
\ee
Eqs.(\ref{huaxieI2}) and  (\ref{I2}),
tell that $M(\tau)$ is the source of
$\beta_0$ and of  $\alpha_2$, simultaneously,
and is expected to contribute equally to them as well.
Eq.(\ref{huaxieI2}) has a formal solution
\be \label{I-2}
\mathcal{I}_2(\tau)  =\int_{0}^{\tau}d\tau'\,
                   M(\tau')e^{-\frac{3}{10}\kappa(\tau,\tau')}
\ee
with $\kappa(\tau,\tau')$ being defined in Eq.(\ref{kappa'}).
As will be seen later, $M(\tau)$ is basically
contributed by the gradient of the peculiar velocity  of photon fluid,
$ k\alpha_1(\tau)$.
When the  perturbation mode $\eta (\tau)$ is specified,
one calculates  $\mathcal{I}_2$ from Eq.(\ref{I-2}) straightforwardly.
Having obtained $\mathcal{I}_1$ and  $\mathcal{I}_2$,
one proceeds to perform the time integrations of
the modes $\alpha_k$ and $\beta_k$.

\begin{center}
{\em\Large 5. Time Integrals for Temperature Anisotropies and
Polarization}
\end{center}

As demonstrated in Appendix,
by projecting $\alpha_k$ and $\beta_k$ on the basis $P_l(\mu)$
in Eqs.(\ref{aT multipole}) and (\ref{aE multipole}),
one obtains
the multipole moments $a_l^T$ and $a_l^E$
for the temperature anisotropies and the electric type of polarization
respectively,
which have been given by Eqs.(\ref{alt}) and (\ref{ale})
as the following
\be \label{aT1}
a_l^{T}= \int_{0}^{\tau_0}d\tau
\left[\left(\dot H+
 \dot H_l \frac{d^{2}}{d\zeta^{2}}\right)e^{-\kappa(\tau)}
  +V(\tau)\left(\mathcal{I}_1++v_b\frac{d}{d\zeta}
  -\frac{3}{4}\mathcal{I}_2(1+\frac{d^{2}}{d\zeta^{2}})\right)\right]
    j_l(\zeta),
\ee
\be  \label{aE1}
a_l^{E}=\frac{3}{4} \left(\frac{(l+2)!}{(l-2)!} \right)^{1/2}
         \int_{0}^{\tau_0}d\tau\
        V(\tau)\mathcal{I}_2(\tau) \frac{j_l(\zeta)}{\zeta^{2}},
\ee
where  the variable $\zeta \equiv k(\tau_0-\tau)$.
These two time integrations have to be carried out.

\begin{figure}
 \resizebox{120mm}{!}{\includegraphics{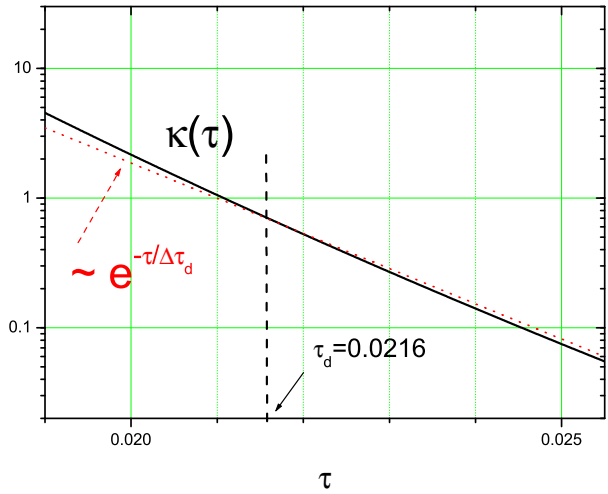}  }
\caption{  \label{fig2} \small The optical depth function
$\kappa(\tau)$ (solid) \cite{HuApJ444}
can approximated by a decreasing
exponential function $\propto e^{-\tau/\Delta\tau_d}$ (dots)
around the recombination.}
\end{figure}

First, we calculate  $a_l^{E}$.
Substituting  $\mathcal I_2$ of Eq.(\ref{I-2})
into Eq.(\ref{aE1}) gives
\be  \label{AlE}
a_l^{E}=\frac{3}{4} \left(\frac{(l+2)!}{(l-2)!} \right)^{1/2}
   \int_{0}^{\tau_0} d\tau V(\tau)\frac{j_l(\zeta)}{\zeta^{2}}
   \int_{0}^{\tau}d\tau'\,
     M(\tau')     e^{-\frac{3}{10}\kappa(\tau')+\frac{3}{10}\kappa(\tau)}.
\ee
As we have seen,
the visibility function $V(\tau)$, as an integrand,  is narrowly peaked around
the recombination time $\tau_d$ with a width $\Delta\tau_d$,
so, for the $\tau$-integration in Eq.(\ref{AlE}),
the integrand around $\tau_d$ only will have significant contributions.
Moreover,
in the $\tau'$-integration
the exponential factor $e^{-\frac{3}{10}\kappa(\tau')}$
behaves like a step function:
$e^{-\frac{3}{10}\kappa(\tau')} \simeq 0$ for $\tau<\tau_d$,
and $e^{-\frac{3}{10}\kappa(\tau')} \simeq 1$ for $\tau >\tau_d$.
Thus, the integrand factor $M(\tau')$ can be approximately pulled
out of the $\tau'$-integration, leading to
\be   \label{pullM}
a_l^{E} = \frac{3}{4} \left(\frac{(l+2)!}{(l-2)!} \right)^{1/2}
    \int_{0}^{\tau_0} d\tau V(\tau) \frac{j_l(\zeta)}{\zeta^{2}}
       M(\tau)
     \int_{0}^{\tau}e^{-\frac{3}{10}\kappa(\tau')+\frac{3}{10}\kappa(\tau)}
      d\tau'.
\ee
To perform  the $\tau'$-integration in Eq.(\ref{pullM}),
one introduces a
variable $x\equiv \frac{\kappa(\tau')}{\kappa(\tau)}$
to replace the variable $\tau'$ \cite{HarariZaldarriaga,Xia2}.
The corresponding limits of integration are
 $\tau'=\tau \rightarrow x=1$ and
 $\tau'=0 \rightarrow x=\infty$.
Since $V(\tau)$ is peaked around $\tau_d$
with a width $\Delta\tau_d$,
one can take $d\tau' \simeq -\frac{dx}{x}\Delta\tau_d$
as an approximation,
valid over the period $\Delta\tau_d$ around the recombination.
For a justification of this approximation,
 in Fig.\ref{fig2}
we plot the optical depth $\kappa(\tau)$
as given in Ref.\cite{HuApJ444},
which indeed behaves approximately as an exponential:
$ \kappa(\tau)\propto e^{-\tau/\Delta\tau_d}$  around $\tau_d$.
Then
\be \label{aE2}
a_l^{E}=
  \frac{3}{4} \left( \frac{(l+2)!}{(l-2)!} \right)^{1/2}\Delta\tau_d
  \int_{0}^{\tau_0}d\tau
     V(\tau)  \frac{j_l(\zeta)}{\zeta^2}
     M(\tau)
   \int_{1}^{\infty}\frac{dx}{x}\
       e^{-\frac{3}{10}\kappa(\tau)x+\frac{3}{10}\kappa(\tau)} .
\ee
$V(\tau)$ contains $  e^{-\gamma (\tau-\tau_d)^2}$,
and  $j_l(\zeta)$ contains a mixture of oscillating modes $e^{ip\tau}$
and  $e^{-ip\tau}$
with $p \propto k$.
Using the formula of a form
$\int^\infty_{-\infty} e^{-\gamma \tau^2} e^{ip\tau}d\tau
= e^{-p^2/4\gamma}\int^\infty_{-\infty}e^{-\gamma \tau^2}d\tau $,
the $\tau$-integration is rendered approximately into
\be
\int_{0}^{\tau_0}d\tau   V(\tau) \frac{j_l(\zeta)}{\zeta^2}  M(\tau)
  \approx
     \frac{j_l(\zeta_d)}{\zeta_d^2}  M(\tau_d) D_E(k)
  \int_{0}^{\tau_0} d\tau    V(\tau),
\ee
where
\be \label{D_E}
D_E(k) = 0.2 (e^{-c_E (k\Delta\tau_1)^{b_E}}
                +e^{-c_E (k\Delta\tau_2)^{b_E}})
\ee
is the
{ Silk damping factor \cite{Silk} for the polarization.
It arises because the CMB photons diffuse through baryons
and smaller scale fluctuations are smoothed, i.e.,
those modes of higher $k$ are more effectively suppressed.
Mathematically, it occurs as
a sort of the Fourier transformation of
 $V(\tau)$ in Eq.(\ref{halfgaussian1}).
This is one of the advantages of our calculation
in that the Silk damping factor arises naturally
instead of adding by hand.}
 $c_E $ and $b_E $ are two fitting parameters,
and the values $c_E\sim 0.27$ and $b_E\sim  2.0$ yield
an agreeing match with
the numerical results \cite{cmbfast,Lewis} over a range $l\lesssim 500$.
The physical interpretation of the appearance of $D(k)$
associated with the recombination process has been given in Refs. \cite{zhao,xia}.
During the recombination around the time $\tau_d$,
the last scattering of CMB photons off baryons
occur effectively only within a time interval $\sim \Delta\tau_d$.
Putting it in terms of the spatial scale,
the smoothing of density fluctuations by the associated
diffusion through baryons
occur effectively only on
a scale of the thickness of the last scattering surface $\sim \Delta\tau_d$
(note that we use unit $c=1$).
Those modes, $e^{ik\tau}$ and $e^{-ik\tau}$,
with wavelengths shorter than $\sim \Delta\tau_d$
are effectively damped,
whereas the long-wavelength modes are less damped.
We remark that, as a fitting formula,
$D_E(k)$ in Eq.(\ref{D_E}) works only approximately,
since other time-dependent factors 
in the integrand have been treated as constants.
Besides, there are other processes  \cite{HuApJ444,BardeenBond}, which 
are significant on small scales, are not taken into account here.
So the parameters $c_E $ and $b_E $ are introduced in Eq.(\ref{D_E}) 
for adjustments.
Among the two terms in Eq.(\ref{D_E}),
$D_E(k)$ is more sensitive to
the term with a smaller time interval $\Delta\tau_1$.

The remaining double integration in Eq.(\ref{aE2})
can be carried straightforwardly
\be
\int_{0}^{\tau_0} d\tau V(\tau)
   \int_{1}^{\infty}\frac{dx}{x}\
       e^{-\frac{3}{10}\kappa(\tau)x+\frac{3}{10}\kappa(\tau)}
=\int^\infty_0 d\kappa  \,  e^{-\frac{7}{10} \kappa}
       \int_{1}^{\infty}\frac{dx}{x}\
       e^{-\frac{3}{10}\kappa x }=\frac{10}{7}\ln\frac{10}{3},
\ee
whereby  Eq.(\ref{V}) has been used in the first equality.
The above treatment of the integrations
is similar to that in Refs.\cite{zhao,xia,Xia2} 
for the case of RGW as the source.
One arrives at the explicit, analytical formula
of the multipole moment of polarization
\be \label{aE3}
a_l^{E} \simeq  \frac{15}{14}\ln\frac{10}{3}
          \left(\frac{(l+2)!}{(l-2)!} \right)^{1/2}
          \frac{ \Delta\tau_d}{k^2(\tau_0-\tau_d)^2}
          M(\tau_d)D_E(k)  j_l(k(\tau_0-\tau_d)),
\ee
which depends upon the function $M(\tau_d)$ at the recombination time $\tau_d$.
As a marked feature,
$a_l^E$ contains explicitly
the recombination width $\Delta\tau_d$,
which arises from the $\tau'$-integration  in Eq.(\ref{pullM}).
Since  $\Delta\tau_d$ is small,
the amplitude of $a_l^E$ will be consequently small,
in comparison with $a_l^T$,
whose dominant part does not contain this $\Delta\tau_d$
as will be seen later {in this Section}.
  Physically, the factor $M(\tau)$ represents the source of
both the leading order polarization $\beta_0$ and
the quadrupole temperature anisotropies $\alpha_2$,
and contributes equally to them  as well.
As its time accumulated effect,
the factor $\Delta {\tau_d} M(\tau_d)$ appears in  $a^E_l$ in Eq.(\ref{aE3})
and in the last term of $a^T_l$ in Eq.(\ref{aT2})
as the quadrupole temperature anisotropies.
During the course of time,
the contribution of $M(\tau)$ is significant
only around the recombination time $\tau_d$ with a width $\Delta {\tau_d}$.
Note that
the spherical Bessel functions
 $ j_l(k(\tau_0-\tau_d))$ in Eq.(\ref{aE3})
is narrowly peaked
around $k(\tau_0-\tau_d)\simeq l$ for $l\gg 1$.
In our notation $\tau_0-\tau_d\sim 0.97$.
So, for each given multipole $l$,
the factor $ j_l(k(\tau_0-\tau_d))$
serves as a filter,
selecting those modes with a wavenumber $k\sim l$ for $a_l^E$.

Next, we calculate  $a_l^T$ in Eq.(\ref{aT1}).
The first term in the integrand of Eq.(\ref{aT1}) is
the integrated Sachs-Wolfe (ISW) contribution,
and contains the exponential factor $e^{-\kappa(\tau)}$,
which can be roughly approximated by a step function \cite{xia,Xia2}:
\be
e^{-\kappa(\tau)}=\left\{
\begin{array}{cc}
0,      &    {\rm for}  ~~ \tau  < \tau_d , \\
1,      &     {\rm for} ~~ \tau_d \le \tau \le \tau_0.
\end{array}
\right.
\ee
So the ISW  term
 is approximated by
\be \label{SachsWolfe}
\int_{0}^{\tau_0}d\tau
\left(\dot H+
 \dot H_l \frac{d^{2}}{d\zeta^{2}}\right)e^{-\kappa(\tau)}j_l(\zeta)
  =\int_{\tau_d}^{\tau_0}d\tau\,
\left( \dot H j_l(\zeta)
     + \dot H_l  \frac{d^2}{d\zeta^2}j_l(\zeta)
   \right),
\ee
where the lower limit has also been replaced by  $\tau_d$.
In the pertinent domain,
$\dot H (\tau)$ and $\dot H_l (\tau)$
 are comparable to each in magnitude,
whereas in Eq.(\ref{SachsWolfe}), the integrated value of
$\frac{d^2}{d\zeta^2} j_l(\zeta)$ is two orders of magnitude
 smaller than
that of $j_l(\zeta)$,
so the term $\frac{d^2}{d\zeta^2} j_l(\zeta)$ in Eq.(\ref{SachsWolfe})
can be neglected in the estimation.
So the left-hand side of Eq.(\ref{SachsWolfe}) reduces to
\be \label{HHL}
 \int_{\tau_d}^{\tau_0}d\tau \dot H j_l(\zeta)
     \simeq  (H(\tau_0)-H(\tau_d))j_l( k\tau_0).
\ee
The second integration in Eq.(\ref{aT1}) is
\be   \label{secondterm}
    \int_{0}^{\tau_0}d\tau\,
 V(\tau)\left[\alpha_0 + \alpha_1 \frac{d}{d\zeta}
-\frac{3}{4} \int_{0}^{\tau}e^{-\frac{3}{10}\kappa(\tau,\tau')}
M(\tau') d\tau'\, \left(1+\frac{d^{2}}{d\zeta^{2}} \right)
\right]j_l(\zeta),
\ee
where  we have substituted $ \mathcal{I}_1 =\alpha_0 $,
$v_b=\alpha_1$ in the tight-coupling limit,
and  $\mathcal{I}_2$ as given in Eq.(\ref{I-2}).
The first two terms in Eq.(\ref{secondterm})
can be treated as before, yielding
\be \label{intalpha0}
\int_{0}^{\tau_0}d\tau\,  V(\tau) \alpha_0(\tau) j_l(\zeta)
        \simeq \alpha_0(\tau_d)
              D_T(k) j_l( k(\tau_0-\tau_d)),
\ee
\be \label{intalpha1}
\int_{0}^{\tau_0}d\tau\,  V(\tau) \alpha_1(\tau) \frac{d}{d\zeta}j_l(\zeta)
          \simeq \alpha_1(\tau_d)
              D_T(k) \frac{d}{d\zeta } j_{l}(  k(\tau_0-\tau_d)),
\ee
where the damping factor for the temperature anisotropies
is taken as
\be
D_T(k)=\frac{1}{2}  (e^{-c_T (k\Delta\tau_1)^{b_T}}
       +e^{-c_T (k\Delta\tau_2)^{b_T}}),
\ee
with $c_T$ and $b_T$ being two fitting parameters,
and   $c_T\sim 0.65$ and $b_T\sim 0.6$ yield
a good match with numerical results  by
CAMB over a range $l\lesssim 500$.
The last term in Eq.(\ref{secondterm})
is a double time integration
and has the same structure as $a_l^E$ in Eq.(\ref{AlE}),
and can be treated in the same way, yielding
\ba
&& -\frac{3}{4}\int_{0}^{\tau_0}d\tau\,  V(\tau)
\left(1+\frac{d^{2}}{d\zeta^{2}} \right) j_l(\zeta)
       \int_{0}^{\tau}e^{-\frac{3}{10}\kappa(\tau,\tau')} M(\tau')
           d\tau'    \nonumber \\
& & \simeq   -\frac{15}{14} \ln\frac{10}{3}
      \Delta\tau_d  M(\tau_d) D_E(k)
          \left( 1+\frac{d^2}{d\zeta^2}  \right)
          j_l(\zeta) |_{\zeta=k(\tau_0-\tau_d)}.
\ea
Again, the term proportional to $\frac{d^2}{d\zeta^2} j_l(\zeta)$ can be neglect
when calculating the power spectrum in Section 7.
Putting these four pieces together,
one arrives at
the explicit, analytical formula
of the multipole moment of temperature anisotropies
\ba  \label{aT2}
a_l^T&& =\alpha_0(\tau_d)  D_T(k)  j_l(k(\tau_0-\tau_d))
     +\alpha_1(\tau_d) D_T(k) \frac{d}{d\zeta } j_{l}(k(\tau_0-\tau_d)
     \nonumber\\
&&+  (H(\tau_0)-H(\tau_d))j_l( k\tau_0)
             \nonumber\\
&&  -\frac{15}{14} \ln\frac{10}{3}
        \Delta\tau_d   M(\tau_d)D_E(k)
        \left( 1+\frac{d^2}{d\zeta^2}  \right)
        j_l(\zeta) |_{\zeta=k(\tau_0-\tau_d)} .
\ea
In the above expression,
the $\alpha_0$ term is dominant,
the $\alpha_1$ term is secondary,
the ISW term is smaller than the  $\alpha_1$ term,
and the last term containing the factor  $\Delta\tau_d$ is smaller than
the ISW term.
Note that the two major terms, $\alpha_0$ and $\alpha_1$,
in Eq.(\ref{aT2}) do not contain $\Delta\tau_d$.
This is because
their corresponding integrations,
Eqs.(\ref{intalpha0}) and (\ref{intalpha1}),
 are single time integrations,
instead of double time integration.
By comparison,
the  amplitude of $a_l^T$ is expected to be
higher than that of $a_l^E$.
We note that the structure of $a_l^T$ in Eq.(\ref{aT2})
is similar to the parallel formula in the Newtonian gauge
given in Ref.\cite{HuApJ444},
which did not have the last term $\propto \Delta\tau_d M(\tau_d)$.
The relative contributions
of the four terms in Eq.(\ref{aT2}) will be demonstrated in Fig.\ref{fig6}.

To completely determine
$a^T_l$  in Eq.(\ref{aE3}) and $a_l^E$ in Eq.(\ref{aT2}),
one still needs $\alpha_0$, $\alpha_1$,
$M(\tau)$ in Eqs.(\ref{alpha0}),   (\ref{alpha1}), and (\ref{M}),
respectively,
which all depend upon the scalar perturbations $h(\tau)$ and $\eta(\tau)$.
In the following we will solve for
$h(\tau)$ and $\eta(\tau)$.

\begin{center}
{\em\Large 6. Determination of  scalar perturbations}
\end{center}

The unperturbed spacetime background  are
described by the Friedmann equations:
\be \label{friedmann1}
(\frac{\dot{a}}{a})^2=\frac{8\pi}{3}Ga^2\bar{\rho},
\ee
\be
\frac{\ddot{a}}{a} = -\frac{4\pi}{3}Ga^2(\bar{\rho}+3\bar{P}),
\ee
where $\bar{\rho}$ and $\bar{P}$ are
the mean energy density and pressure.
The Einstein equations for the scalar perturbations
in synchronous gauge are the following \cite{Grishchuk density,ma}
 \be\label{pert-eq1}
 k^2 \eta  - \frac{1}{2}\frac{{\dot a}}{a}\dot h
            = 4\pi Ga^2 \delta T_0^0 ,
\ee
\be \label{pert-eq2}
  k^2 \dot \eta
    = 4\pi Ga^2 ik^j\delta T^0_j  ,
\ee
\be \label{pert-eq3}
 \ddot h + 2\frac{{\dot a}}{a}\dot h - 2k^2 \eta
        =  - 8\pi Ga^2\delta T_i^i ,
\ee
\be  \label{pert-eq4} \ddot h + 6\ddot \eta  + 2\frac{{\dot
a}}{a}\left( {\dot h + 6\dot \eta } \right) - 2k^2 \eta
     =  - 24\pi Ga^2 (\bar \rho  + \bar P)\sigma ,
\ee
where $\sigma$ represents the anisotropic stress,
$\delta T^0_0=-\bar{\rho}\delta $ is
the perturbed energy density,
 $\delta T^i_i=3\delta P$ is the perturbed  pressure,
and $\frac{\delta P}{\delta \rho}=c_s^2$,
where the sound speed
$c_s \simeq \frac{1}{3}$ in  RD era
and $c_s \simeq 0$ in  MD era.

First, let us do for  the RD era.
We are concerned with the long wave modes with $k\tau \ll 1$,
and the solutions of the set of Eqs.(\ref{pert-eq1})-(\ref{pert-eq4})
for  $h$ and $\eta$ are  \cite{Press,Ratra,Grishchuk density,ma}
\ba \label{h}
&&h= A + B(k\tau)^{-2}+C(k\tau)^2+D(k \tau),\\
&& \eta=2C +\frac{3}{4}D(k\tau)^{-1}. \label{eta1}
\ea
All the coefficients $A$ through $ D$
actually depend on the comoving wavenumber $k$,
which has been skipped hereafter for notational simplicity.
The two terms proportional to $A$ and $B$ are gauge modes,
which will be  dropped,
and  two physical modes are proportional to $C$ and $D$.
\be  \label{hr}
h=C(k\tau)^2+D(k\tau).
\ee
Among these two modes,
the mode $(k\tau)^2$ grows faster and is dominant at late times,
and the  mode $(k\tau)$ is less important,
which was neglected in the treatment of Ref.\cite{ma}
and was taken to be small in Ref.\cite{Grishchuk density}.
In principle, the two coefficients $C$ and $D$ should
be determined by either the inflationary  or the reheating era
that precedes the RD era.
To avoid further complication from the preceding eras,
we shall treat $D$ as a small parameter proportional to $C$.
For simplicity of analytical calculations,
we do not include the modifications due to cosmic neutrinos,
which will bring higher order terms $(k\tau)^2$
to $\eta$ during the RD era \cite{ma}.
Let us examine the long wave approximation during  the RD era.
At the radiation-matter equality $\tau_2$
the comoving sound horizon is $\sim c_s  \tau_{2}$.
Those $k$-modes with $1/k > c_s  \tau_{2} $
can be taken as the long wave modes during the RD era.
In our notation with the comoving time $\tau$
specified from Eq.(\ref{metric}) through Eq.(\ref{a2}),
this is equivalent to $k \lesssim 210$ (0.025 Mpc$^{-1}$).
For wave number greater than  this,
a more elaborated treatment of the perturbations during the RD
would be desired than presented here.

For the MD era, the solution of the metric perturbations are given by
 \cite{Press,Ratra,Grishchuk density,ma}
\be   \label{h2}
h=J +(k\tau) ^2E+\frac{1}{(k\tau)}F + \frac{1}{(k\tau)^3}G,
\ee
\be  \label{eta join}
\eta= 5E-\frac{1}{(k\tau)^3}F,
\ee
 where the constant  $J$ is a gauge mode
 corresponding to
 a transformation of the spatial coordinates,
 i.e., a rescaling of the scale factor $a(\tau)$,
 and can be dropped.
As has been known \cite{Press},
for $h(\tau)$ in Eq.(\ref{h2}),
the linear combination
$\frac{2}{(k\tau)} + \frac{1}{(k\tau)^3}$
is another gauge mode,
which is dominated by $\frac{1}{(k\tau)^3}$ for $k\tau\ll 1$
(long wavelengths or early time),
and by   $\frac{1}{(k\tau)}$ for $k\tau\gg 1$
(short wavelengths or late time).
In our context,
we aim at the large angular temperature anisotropies and polarization
of CMB.
So we are concerned with the long wavelength perturbations
around the radiation-matter equality $\tau_2$
and the recombination time $\tau_d$.
Thus the $G$ term is taken as the dominant gauge mode.
To keep our analytical calculation simple,
we drop the $G$ term.
In fact, the $G$ term is the time-translation-invariant solution
and can be gauged away by a restricted coordinate transformation
within in the synchronous gauge \cite{Ratra}.
Other discussions on gauge modes are given
in Refs. \cite{Press,Grishchuk density}.
The term proportional to $E$ in Eq.(\ref{h2}) grows with time
and is the primary portion of the physical mode.
Thus, for the MD era, one has
\be \label{hm}
h= (k\tau) ^2E+\frac{1}{(k\tau)}F,
\ee
\be \label{eta}
\eta=5E-\frac{1}{(k\tau)^3}F .
\ee
From these specifications,
the source $S(\tau)$ in Eq.(\ref{S}) reduces to
\be \label{SRD}
S(\tau)= \frac{2 k^2}{3 }C  ,\,\,\,\,\,\, {\rm for \,\,\, RD},
\ee
\be \label{SMD}
S(\tau) = \frac{2k^2}{3}E+\frac{2}{3k\tau^3}F  , \,\,\,\,\,\, \rm for \,\,\, MD.
\ee

Now we need to make a proper connection of the perturbations
for the RD and MD eras at the radiation-matter equality $\tau=\tau_2$.
We remark  that
the energy density perturbation $\delta $ is continuous
in the transition from  RD to MD era.
But the pressure $P$ is not required to be so,
as
$P>0$ during RD, and $P=0$ during MD.
By the perturbed Einstein equation Eq.(\ref{pert-eq1}) for $\delta $,
one finds that
the combination $ k^2 \eta  - \frac{1}{2}\frac{{\dot a}}{a}\dot h$
is required to be continuous  at $\tau=\tau_2$.
Since  $\dot a(\tau)$
is continuous as prescribed in Eqs.(\ref{a1}) and (\ref{a2}),
$\dot h$ and $\eta$ are  required to  be continuous,
leading to
\be
2 k^2\tau_2 C+kD =2k^2\tau E-\frac{F}{k\tau^2_2}\, ,
\ee
\be
2C +\frac{3}{4k\tau_{2}}D
                      =5E-\frac{1}{(k\tau_{2})^3}F \, .
\ee
From these two algebraic equations,
one solves for $E$ and $F$ in terms of  and $C$ and $ D$.
\ba
&&E=-\frac{D}{12 k\tau_2}\, , \\
&&F=-(k\tau_2)^3 \left( 2C +\frac{7D}{6k\tau_2} \right)\, .
\ea
The coefficient $D$  is a small parameter that needs to be fixed.
We take the coefficient $D$ to be smaller than $C$ by a factor $(k\tau_2)$
in the long wavelength limit $k\tau\ll 1$,
so that $D\sim (k\tau_2)C$.
Specifically, in the following analytical calculations,
we take
\be \label{D}
D=-\frac{24}{5}(k\tau_2)C,
\ee
though
other possible choices may also be justified as long as
$D$ is subdominant to $C$ in the long wavelength limit.
Substituting Eq.(\ref{D}) into the above yields
\be \label{FC}
       E=\frac{2}{5}C, ~~~\,\,\,\,  F=\frac{18}{5}(k\tau_2)^3C.
\ee
 One can check that, in the RD,  as well as in MD era,
if we transform the perturbations $h$ and $\eta$ in synchronous gauge
back to the $\phi$ and $\psi$ in Newtonian gauge \cite{HuApJ444},
the results are consistent with each other.
Fig. \ref{fig4} shows
the continuous joining of
the perturbation modes $h(\tau)$ and $\eta(\tau)$
at $\tau_2$,
and Fig. \ref{fig5} shows the continuous joining of
the modes $H(\tau)$ and $H_l(\tau)$.
As one can check,
the functions  $\ddot h$ and $\dot \eta$
are not continuous at $\tau=\tau_2$ by our joining condition.

To fix the initial condition,
we  need to specify the $k$-dependent coefficient $C$.
According to the inflationary models of the early universe,
the primordial scalar perturbations were generated
with a nearly scale-invariant spectrum
with a spectral index  $n_s\sim 1$ \cite{Guth}.
In our notation
this corresponds to $C \propto k^{\frac{1}{2}(n_s-3)}$.
For inflationary models proposed so far,
the most uncertain quantity is the amplitude of the spectrum.
In practice, this can be fixed by cosmological observations,
say, the WMAP result.
One writes the curvature perturbation spectrum
\be \label{DeltaR}
\Delta^2_{R}(k) = \Delta^2_{R}(k_0)
    \left(\frac{k}{k_0}\right)^{n_s-1+\frac{1}{2}\alpha_s \ln (k/k_0)},
\ee
where the physical pivot wavenumber $k_0=0.002$Mpc$^{-1}$,
$\Delta^2_{R}(k_0)$ is the normalization at $k_0$.
WMAP5 \cite{WMAP5} gives
$\Delta^2_R(k_0)=(2.41\pm 0.11)\times10^{-9}$,
WMAP5+BAO+SN Mean \cite{Komatsu} gives
$\Delta^2_R(k_0)=(2.445\pm0.096)\times10^{-9}$.
Besides the scalar spectral index $n_s$,
we include a possible scalar running spectral index $\alpha_s$
in the spectrum \cite{Liddle,Kosowskyturner}.
The fitted value of $n_s$ is much affected
by the presence of $\alpha_s$ and the RGW component,
and by additional combination with SN Ia and BAO data as well.
In absence of  $\alpha_s$ and the RGW,
WMAP5 gives $n_s=0.963^{+0.014}_{-0.015}$ \cite{WMAP5},
WMAP5 +SN Ia+BAO gives  $n_s=0.960^{+0.014}_{-0.013}$ \cite{WMAP5},
and
WMAP7 gives $n_s=0.963 {\pm 0.012}$ \cite{Komatsu}.
When $\alpha_s$ is allowed,
WMAP5 gives $n_s=1.087^{+0.072}_{-0.073}$
           and $\alpha_s=-0.050 {\pm 0.034}$
           with a better determination of the third acoustic peak   \cite{Komatsu},
WMAP5+BAO+SN has given $n_s=1.089^{+0.070}_{-0.068}$
           and $\alpha_s=-0.053^{+0.027}_{-0.028}$ \cite{WMAP5,Komatsu}.
More recent
%WMAP7 data-only gives $n_s=1.027^{+0.020}_{-0.051}$
    %       and $\alpha_s=-0.034\pm 0.026$ \cite{WMAP7}, and
WMAP7+ACBAR+QUaD gives $n_s=1.041^{+0.045}_{-0.046}$
           and $\alpha_s=-0.041^{+0.022}_{-0.023}$ \cite{WMAP7,Komatsu}.
When RGW is also allowed \cite{TongZhang},
WMAP7+Tensor gives $n_s=1.076\pm 0.065$,
                   $\alpha_s=-0.048\pm 0.029$$, r<0.49$ \cite{Komatsu}.
In the slow-roll scalar inflationary models,
$n_s$  and $\alpha_s$ can be calculated
from the inflationary potential
and its derivatives \cite{Liddle,Kosowskyturner}.
For generality,
we will treat $n_s$ and $\alpha_s$ as parameters.
Eq.(\ref{DeltaR}) corresponds to
\be \label{C}
 C = C_0
    \left(\frac{k}{k_{C}}\right)^{\frac{1}{2}(n_s-3)
    +\frac{1}{4}\alpha_s \ln (k/k_0)},
\ee
where the normalization $C_0 \sim 204$.
The physical pivot wavenumber  $k_{0}$
corresponds to a comoving wavenumber
$k_{C}=k_0 a(\tau_0) \simeq 17.1$
for the Hubble parameter $H_0=70.1$ km s$^{-1}$ Mpc$^{-1}$ \cite{Komatsu}.

In fixing the initial condition for
$\alpha_0$ at $k\tau\ll 1$ during the RD era,
the coefficients $B_1$ and $ B_2$ in Eq.(\ref{alpha0})
have to be specified.
In the tight-coupling approximation,
 $k\alpha_1$ in  Eqs.(\ref{alpha_0}) can be neglected,
 yielding  $\alpha_0=h/3$ for  $k\tau\ll 1$.
By comparison, in the limit $k\tau\rightarrow 0$,
$h$ behaves as in Eq.(\ref{hr}),
so the term $B_1\cos(p\tau)$ in $\alpha_0$ should be vanishing,
leading to  $B_1=0$.
The  $B_2\sin(p\tau)$ term in Eq.(\ref{alpha0})
represents the isocurvature mode of initial perturbations.
A stringent constraint has been given by WMAP5
on the isocurvature contribution with
the isocurvature/adiabatic ratio
$\alpha_{-1} < 0.015$ at $95\%$ CL  \cite{WMAP5,Xia2}.
For simplicity,
we can choose the coefficient $B_2=0$.
Then the monopole $\alpha_0(\tau)$ in Eq.({\ref{alpha0}})
reduces to the integration
\be \label{alpha00}
\alpha_0(\tau)   =
\int_0^{\tau_2}\frac{S(\tau')\sin{(p\tau-p\tau')}}{p}d\tau'
   +\int_{\tau_2}^{\tau}\frac{S(\tau')\sin{(p\tau-p\tau')}}{p}d\tau',
\ee
Using
Eqs.(\ref{SRD}), (\ref{SMD}),  (\ref{FC}) into the above yields
the monopole
\ba \label{0}
\alpha_0(\tau)
&&= 2 C(1+R)
   \left[ \frac{3}{5}\cos(p\tau-p\tau_2)
    -\cos{p\tau}+ \frac{2}{5}  \right] \nonumber\\
&&   +C (1+R) \frac{36}{5}
          \int_{\tau_2}^{\tau}\frac{1}{k^3\tau\,'^3 }
              \sin{(p\tau-p\tau')}d(p\tau').
\ea
From Eq.(\ref{alpha1}) and Eq.(\ref{hm}) follows the dipole
\ba \label{1}
\alpha_1(\tau)
&&=  2C\sqrt{3(1+R)}
           \left[ \frac{3}{5} \sin (p\tau-p\tau_2) -\sin(p\tau) \right]
                              \nonumber \\
  &&  -C\frac{36}{5} \sqrt{3(1+R)}
    \int_{\tau_2}^{\tau}  (\frac{\tau_2}{\tau\,' })^3
       \cos (p\tau-p\tau') \,  d(p\tau') \nonumber \\
 &&      +C \frac{4}{5} (k\tau)
       -C\frac{18}{5} (\frac{\tau_2}{\tau})^2 (k\tau_2).
\ea
By the definition in Eq.(\ref{M}),
we take  time derivatives of $\alpha_0(\tau)$ and $\eta(\tau)$
in Eq.(\ref{0}) and Eq.(\ref{eta join}), respectively,
and arrive at
\ba \label{Mf}
M(\tau)
  &=&-\frac{4}{5}C(1+R)
  \left[ -\frac{3}{5}p\sin (p\tau-p\tau_2)
                    +p\sin p\tau  \right] \nonumber \\
&-&  \frac{72}{25}C(1+R)
    \int^{\tau}_{\tau_2}
    \frac{\tau_2^3 p }{\tau'^3}\cos(p\tau  -p\tau')d(p\tau')
    -  C\frac{216}{25}\frac{\tau_2^3}{\tau^4}
\ea
for the MD era.
We have checked that, in this final expression,
$M(\tau)$ is dominated by
the first two terms coming from $-2 \dot \alpha_0(\tau)/5$,
whereas the last term $-C\frac{216}{25}\frac{\tau_2^3}{\tau^4}$
coming from $-4\dot \eta(\tau) /5 $ is comparatively small
by more one order of magnitude.
It is important to notice that,  due to time differentiation,
$M(\tau)$ contains the functions like $\sin (p\tau-p\tau_2)$,
whereas $\alpha_0(\tau)\propto \cos (p\tau-p\tau_2)+...$.
This fact will lead to the character of the present CMB
that
the peaks of the polarization
$C^{EE}_l \propto |a^E_l|^2 \propto |\sin(p\tau_d-p\tau_2)|^2+...$
and of the temperature anisotropies
$C^{TT}_l \propto |a^T_l|^2 \propto |\cos(p\tau_d-p\tau_2)|^2+...$
appear alternatingly.

Using  Eqs.(\ref{relation}),   (\ref{hm}),  (\ref{eta}), and (\ref{FC}),
the time derivatives of the  scalar modes,
$\dot H$ and $\dot H_l$ during MD, are given by
\be  \label{H}
\dot H(\tau) =C\frac{108}{5}\frac{\tau_2^3}{\tau^4},
\ee
\be \label{Hl}
\dot H_l (\tau) = -\frac{2}{5}C
\left(2k^2\tau -9k^2\frac{\tau_2^3}{\tau^2}
 +162 \frac{\tau_2^3}{\tau^4} \right).
\ee
So $\dot H (\tau)$ and $\dot H_l (\tau)$
 are comparable to each in magnitude around the recombination.

\begin{center}
{\em\Large 7. The analytical spectra }
\end{center}

Given the multipole moments
$a^E_l$ in Eq.(\ref{aE3}) and  $a^T_l$ in Eq.(\ref{aT2}),
the spectra $C_l^{TT}$, $C_l^{TE}$, and $C_l^{EE}$ are calculated
as the following integrations over the wavenumber $k$ \cite{Baskaran}
\ba
&&C_l^{TT}     = \int  |a^T_l(k)|^2 kdk , \\
&&C_l^{TE}     =\int  a^T_l(k)a^E_l(k) kdk  , \\
&&C_l^{EE}     =\int |a^E_l(k)|^2  kdk.
\ea
The resulting spectra are explained in the following graphs.

\begin{figure}
 \resizebox{120mm}{!}{\includegraphics{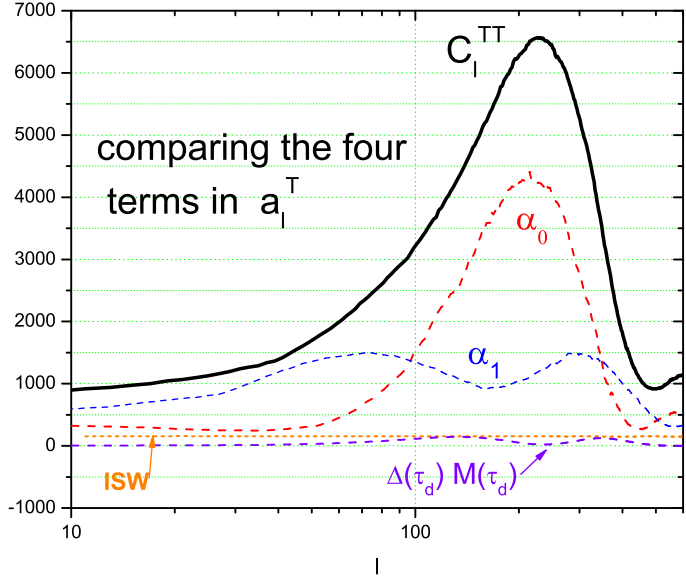}  }
\caption{  \label{fig3} \small
The multipole moment $a_l^T$ has four terms
in Eq.(\ref{aT2}),
which are schematically plotted for a comparison.
$\alpha_0$ is dominant at $l\sim 200$,
$\alpha_1$ is dominant at $l\lesssim 100$,
the ISW is flat and low, and
$\Delta\tau_d M(\tau_d)$ term is the lowest with two small bumps.
}
\end{figure}

\begin{figure}
 \resizebox{140mm}{!}{\includegraphics{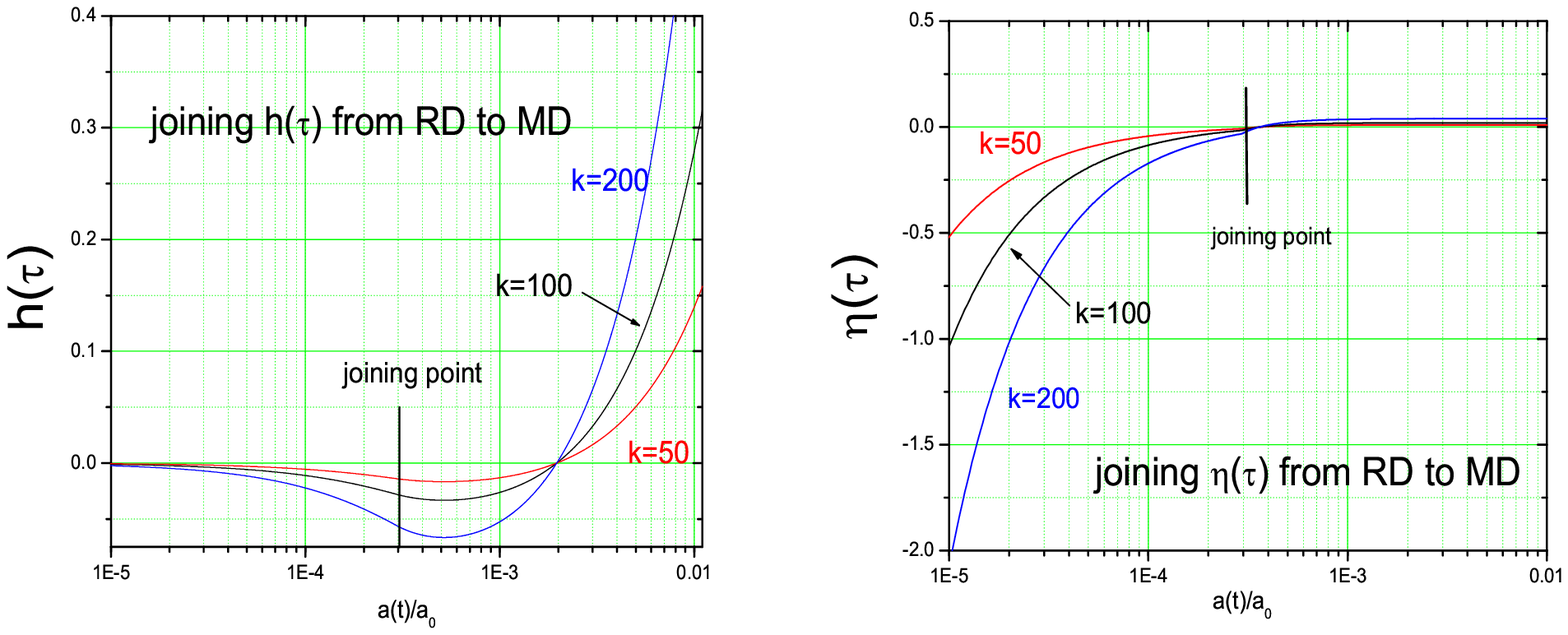}  }
\caption{  \label{fig4} \small
The perturbation modes $h(\tau)$ and $\eta(\tau)$
        continuously joined at $\tau_2$,
        respectively.}
\end{figure}

\begin{figure}
 \resizebox{140mm}{!}{\includegraphics{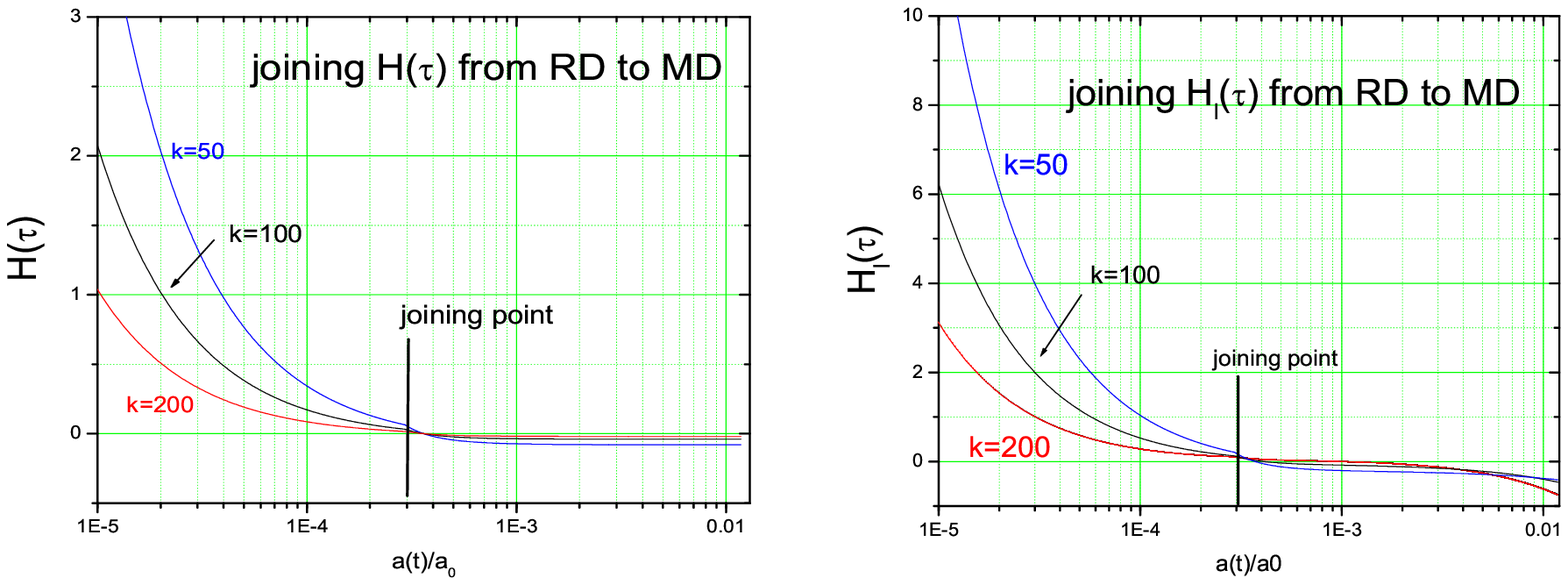}  }
\caption{  \label{fig5} \small
The perturbation modes $H(\tau)$ and $H_l(\tau)$
        continuously joined at $\tau_2$,
        respectively.}
\end{figure}

\begin{figure}
 \resizebox{120mm}{!}{\includegraphics{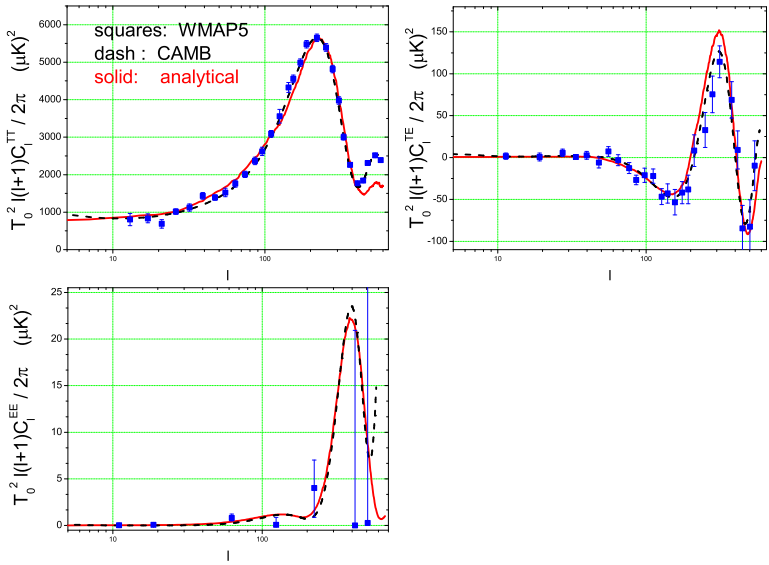}  }
\caption{  \label{fig6} \small
The analytical spectra (red line) $T_0^2 l(l+1)C_l^{XX'} /2\pi$
compared with the numerical result (dash lines) of CAMB \cite{Lewis}
and the observed (square dots) WMAP5 \cite{WMAP5}.
}
\end{figure}

Fig. \ref{fig3} demonstrates the relative contributions by each term
to $C_l^{TT}$.
Over the relevant range $l \lesssim  500$,
the contribution by ISW is rather flat as a function of $l$,
and its amplitude is at most
  $ \sim 10\%$ that of the  $\alpha_1$ term.
The last term in Eq.(\ref{aT2}) contains
the factor $\Delta\tau_d M(\tau_d)$
given by setting $\tau=\tau_d$ in  Eq.(\ref{Mf}),
and its  contribution  to $a_l^{T}$ is even smaller than the ISW term,
with two low bumps at $l\sim 130$ and at $l\sim 350$.
The smallness of this term is due to the extra small
factor $\Delta\tau_d \sim 0.003$.
Thus, the major features of $a_{l}^T $ in  Eq.(\ref{aT2})
are largely contributed by
$\alpha_0(\tau_d)$ and  $ \alpha_1(\tau_d)$,
whereas the quadrupole part of $a_l^T$
contains the factor $\Delta \tau_d M(\tau_d)$,
similar to the polarization $a_l^E$ in Eq.(\ref{aE3}).
This also tells that the polarization $a^E_l$ is smaller
than the temperature anisotropies $a^T_l$ in amplitude.

\begin{figure}
 \resizebox{120mm}{!}{\includegraphics{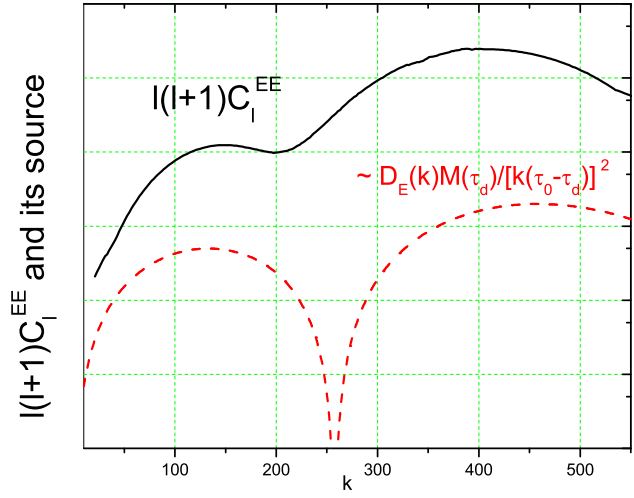}  }
\caption{  \label{fig7} \small
The profile of $l(l+1)C_l^{EE}$ (black line) is determined by
its source $ D_E(k) M(\tau_d)/(k(\tau_0-\tau_d))^2$
in Eq.(\ref{aE3}).
In particular, the bump locations of $l(l+1)C_l^{EE}$
is determined by that of the scalar perturbations (red dotted line).
}
\end{figure}

In Fig.\ref{fig6}
we plot these analytical spectra $l(l+1)C_l^{TT}$,
$l(l+1)C_l^{TE}$, $l(l+1)C_l^{EE}$
for  $n_s=0.96$, $\alpha_s=0$,
and the baryon fraction $\Omega_b=0.045$.
For comparison, the numerical result from CAMB \cite{Lewis}
and the observed result from WMAP5 \cite{WMAP5}
are also given.
For a more realistic case,
one would have to also include
the analytical $C_l^{XX'}$ by RGW \cite{zhao,xia,Xia2}
at a tensor/scalar ratio $r$
to form the complete calculated $C_l^{XX'}$.
We leave that for future studies.
Fig. \ref{fig7} shows that the overall profiles of
the analytical spectra agree well with the numerical
and the observed
on large angular scales with $l\lesssim 500$.
This range covers the first primary peak of $C_l^{TT}$ and
the first two primary peaks of $C_l^{EE}$ and of $C_l^{TE}$.
Only around $l\simeq 310$ where the second primary peak of $C_l^{TE}$
is located,
the analytical $C_l^{TE}$ deviates by $\sim 18\%$ higher in amplitude
from the numerical one.
For smaller angular scales,
the analytical results deviate considerably from the numerical ones.
This has been expected since our calculation
is based upon the long wavelength approximation
valid only for large angular scales.
From  Fig.\ref{fig6},
we see that the first two peaks of  $C_l^{TT}$ occur
at $l\sim 200$ and $l\sim 500$,
while those of  $C_l^{EE}$ occur
at $l\sim  100$ and $l\sim 400$.
This alternating occurrence of the peak locations of $C_l^{EE}$ and  $C_l^{EE}$
has been anticipated.
(See  the discussion below Eq.(\ref{Mf}).
Based on the analytic results,
one can estimate the span  of the two adjacent peaks of $C^{EE}_l$ in $l-$ space,
which corresponds to that of $|\sin (c_s k (\tau_d - \tau_2) )|^2$ in $k-$ space.
Since
$ j_l(k(\tau_0-\tau_d))$
is significantly contributive only
around $k(\tau_0-\tau_d) \sim l$ for $l\gg 1$,
it  plays a role of a filter and selects
those $k(\tau_0-\tau_d) \sim l$ part of the integrand
to contribute to the integration $\int dk$ over $k$.
Qualitatively,
the span $\Delta k$ of two adjacent peaks of $|\sin (c_s k (\tau_d - \tau_2) )|^2$
is given by a relation $\pi = c_s \Delta k (\tau_d - \tau_2)$.
Then the span of the two adjacent peaks of $C^{EE}_l$
in $l-$ space is $ \Delta  l\sim  \Delta k (\tau_0-\tau_d) \sim 370$.
The same $ \Delta  l$ holds also for $C^{TT}_l$.
This is roughly what is seen in Fig.\ref{fig7}.
(See also Ref.\cite{montanari}).

In Fig.\ref{fig7},
we sketch the  profile of  $l(l+1)C_l^{EE}$ as a function of $l$,
which, notably, has two bumps,
one at $l\sim 100$, and another at $l\sim 400$.
In order to interpret the origin of these two bumps,
we also sketch the main factor $D_E(k) M(\tau_d)/[k(\tau_0-\tau_d)]^2$
of $a^E_l$ in Eq.(\ref{aE3})  as a function of $k$.
By  the projection of $ j_l(k(\tau_0-\tau_d))$,
the square of
$ D_E(k) M(\tau_d)/[k(\tau_0-\tau_d)]^2$ around $k\sim l$, aside some factor,
is basically $C_l^{EE}$ around $l$.
Since $ D_E(k) M(\tau_d)/[k(\tau_0-\tau_d)]^2$ has two bumps,
around $k\sim 100$ and $k\sim 400$,
they give rise to the two bumps of  $C_l^{EE}$.

\begin{figure}
 \resizebox{120mm}{!}{\includegraphics{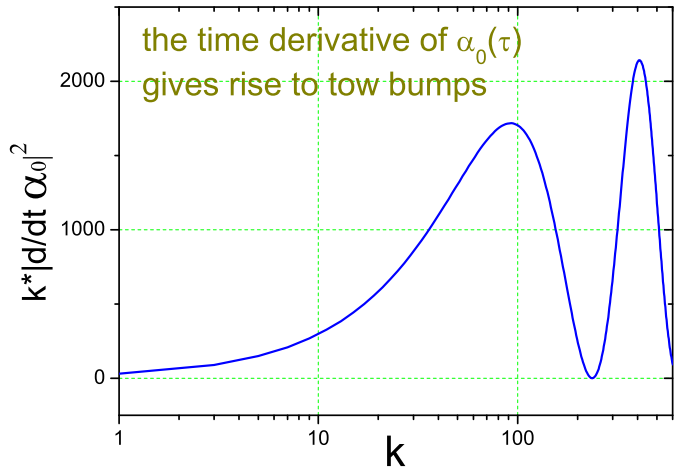}  }
\caption{  \label{fig8} \small
It is the time derivative $\dot \alpha_0(\tau)$
that gives rise to the first two bumps in $C_l^{EE}$.
}
\end{figure}

Fig.\ref{fig8} shows the first two peaks of
the squared time derivative $k|\dot \alpha_0(\tau)|^2$.
Below Eq.(\ref{Mf}) we have mentioned that $-2 \dot \alpha_0(\tau)/5$
is the dominant term of $M(\tau)$.
Thus, it is the time derivative $\dot \alpha_0(\tau)$
that essentially determines the characteristic profile  of $C_l^{EE}$,
including the peak locations.
The two peaks of $|\dot \alpha_0(\tau)|^2$
consequently gives rise to the first two peaks of $C_l^{EE}$,
the first one actually being very low so that it is only a low bump.

\begin{figure}
 \resizebox{120mm}{!}{\includegraphics{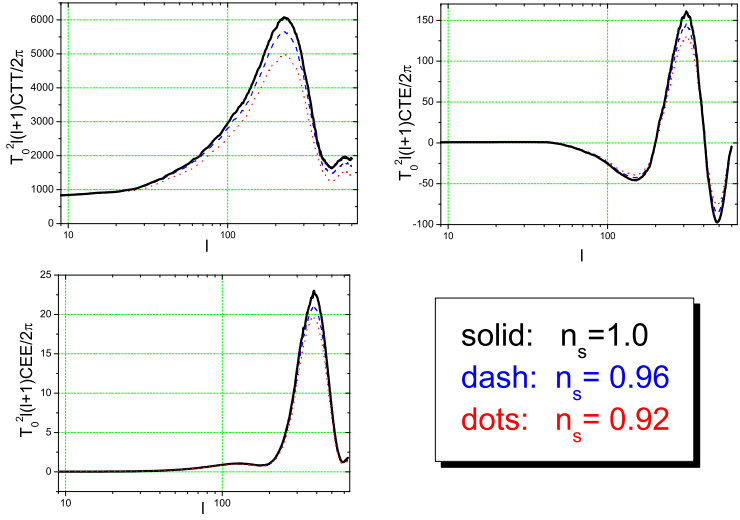}  }
\caption{  \label{fig9} \small
 The spectra $C_l^{XX'} $ depend on the  primordial power index $n_s$
 of the scalar metric perturbations.
 }
\end{figure}

Fig.\ref{fig9} shows
the dependence of  $C_l^{XX'} $
upon the scalar spectral index $n_s$.
The pivot point $k_0=0.002$Mpc$^{-1}$
corresponds to $l\sim 12$.
As is seen,
a greater value of $n_s$ yields a higher amplitude of $C_l^{XX'} $
for $l > 12$.
This is expected
from the initial amplitude $C$ given in Eq.(\ref{C}),
which gets larger for a  greater $n_s$ in the range $k\geq k_{0C}$.
The effect  is most obvious around the primary peaks.

\begin{figure}
 \resizebox{120mm}{!}{\includegraphics{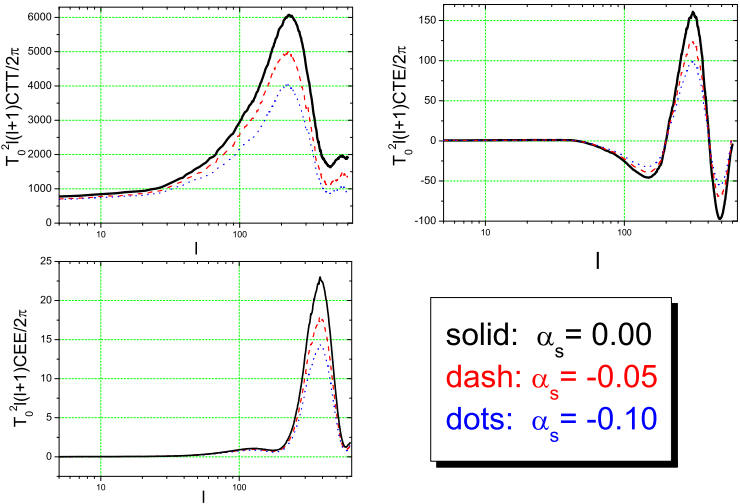}  }
\caption{  \label{fig10} \small
 $C_l^{XX'} $ depend on the running index $\alpha_s$
  of the scalar metric perturbations.
 }
\end{figure}

Fig.\ref{fig10} shows
the dependence of $C_l^{XX'} $
upon the scalar running spectral index $\alpha_s$.
A greater $\alpha_s$ yields a higher amplitude of $C_l^{XX'} $
as is expected from Eq.(\ref{C}) in the range $k\geq k_{0C}$.
Comparing Fig.\ref{fig9} with Fig.\ref{fig10} reveals that
there is a certain degree of degeneracy between
the indices $n_s$ and $\alpha_s$
as two major cosmological parameters.
This degeneracy has demonstrated itself in fitting the observational data of
WMAP \cite{WMAP3,WMAP5,WMAP7,Komatsu}.
Therefore, given the accuracy of current observational data of $C_l^{XX'} $,
it is not easy to distinguish the fine details of
the inflation potentials.

\begin{figure}
 \resizebox{120mm}{!}{\includegraphics{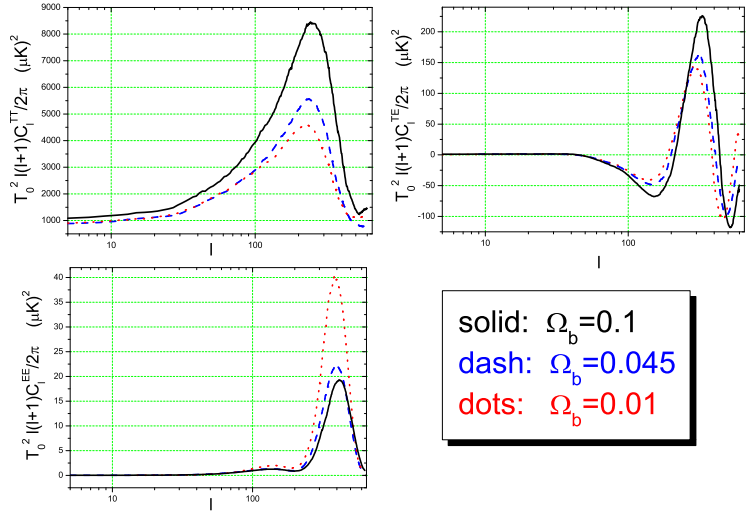}  }
\caption{  \label{fig11} \small
 $C_l^{XX'} $ depend on the baryon fraction $\Omega_b$.
 }
\end{figure}

Fig.\ref{fig11} shows
the dependence of  $C_l^{XX'} $
upon the baryon fraction $\Omega_b$,
in the amplitudes and the locations of peaks and troughs.
As is seen,
a greater value of $\Omega_b$ yields higher amplitudes of $C_l^{TT} $
({also see Refs. \cite{HuApJ444},\cite{peebles}, \cite{peebles2}, \cite{jones-wise}})
and  $C_l^{TE} $, but a lower amplitude of $C_l^{EE} $.
This can be understood as follows.
 Eqs.(\ref{0}) and (\ref{1}) show that a greater $\Omega_b$
corresponds to a greater $R$ and
gives higher amplitudes of $\alpha_0$ and $\alpha_1$,
hence a higher amplitude of $a_l^T$ in Eq.(\ref{aT2})
and  of $C_l^{TT} $.
On the other hand,
$a_l^E$ in Eq.(\ref{aE3}) is proportional to
the recombination width $\Delta\tau_d$,
which is smaller for a greater value of $\Omega_b$
as fitted by Eq.(\ref{Deltatau}).
Thus $C_l^{EE} $ has a lower amplitude for a greater $\Omega_b$.
As for the cross spectrum $C_l^{TE} $,
the $\Omega_b$-dependence of its amplitude
is the outcome of these two competing factors.
Since greater $n_s$ and $\Omega_b$ both tend to
enhance the amplitudes of the spectra $C_l^{TT} $ and $C_l^{TE} $,
there is also a degeneracy between
 $n_s$ and $\Omega_b$ in regards to $C_l^{TT} $ and $C_l^{TE} $.
Nevertheless,
for the spectrum $C_l^{EE}$,
greater $n_s$ and $\Omega_b$ have just opposite
effects on its amplitude.
This feature will help to
break the degeneracy.
Fig.\ref{fig11} also shows that
a greater  $\Omega_b$ shifts
the locations of peaks and troughs of  $C_l^{XX'}$
to larger $l$ (smaller angles).
This is because
a greater $\Omega_b$ leads to a lower sound speed $c_s$
of photon gas,
so at a fixed frequency
the corresponding wavelength is suppressed \cite{Naselsky}.
By the analytic results,
this is evident from
the oscillating factors $\sin(c_s k\tau_0)$ and $\cos(c_s k\tau_0)$
contained in $a_l^T$ and $a_l^E$,
whose peak locations are stretched to a larger wavenumber $k$
( i.e., larger $l$ via the projection of $j_l(k(\tau_0-\tau_d))$ )
for a smaller $c_s$.

\begin{figure}
 \resizebox{120mm}{!}{\includegraphics{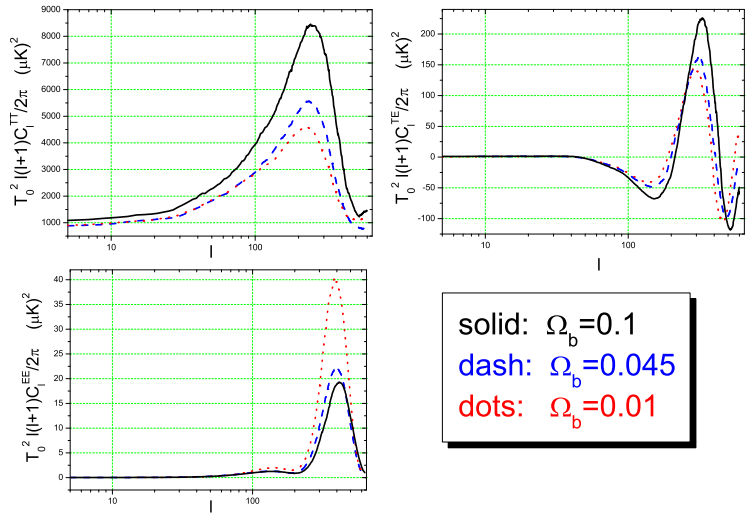}  }
\caption{  \label{fig12} \small
As the analytic expression tells,
a longer recombination process (greater $\Delta\tau_d$)
yields a higher amplitude of polarization $C_l^{EE} $,
and brings  more small-scale damping.
 }
\end{figure}

Fig.\ref{fig12} shows that
a longer recombination process (a greater $\Delta \tau_d$)
yields a higher amplitude of polarization.
This property has been  obvious
since the analytic expression in Eq.(\ref{aE3})
tells  $a_l^{E} \propto \Delta\tau_d$.
Fig.\ref{fig12} also shows  that
a longer recombination process
brings  more damping of  $C_l^{EE}$  on small scales.
This is because $a_l^E$ in Eq.(\ref{aE3})
contains the damping factor $D_E(k)\propto e^{-c_E(k\Delta \tau_d)^2}$.
Similarly, this feature also is shared by $C_l^{TT}$,
as the damping factor $D_E(k)\propto e^{-c_T(k\Delta \tau_d)^2}$
appears in the major term of $a_l^T$ in Eq.(\ref{aT2}).

\begin{figure}
 \resizebox{120mm}{!}{\includegraphics{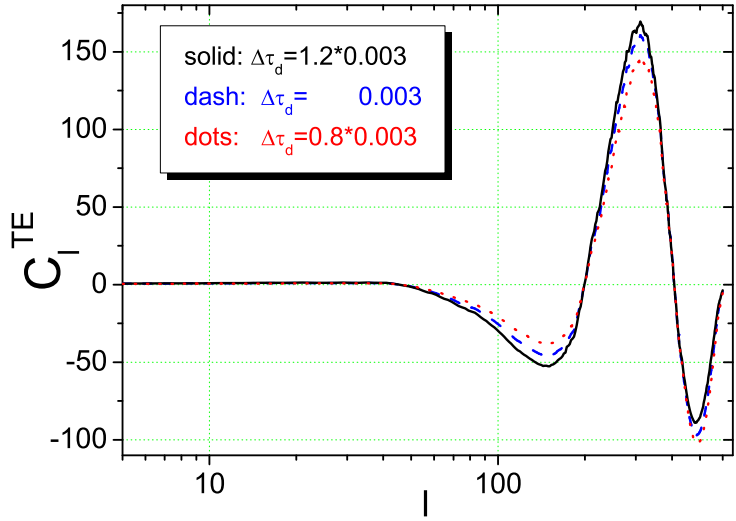}  }
\caption{  \label{fig13} \small
A longer recombination process
yields higher peak and lower trough of cross-correlation $C_l^{TE} $.
 }
\end{figure}

{
Fig.\ref{fig13} shows  that
a longer $\Delta \tau_{d}$ yields
higher peaks as well as lower troughs of $C_l^{TE}$.
Moreover,
a longer $\Delta \tau_{d}$ slightly shifts 
the peaks and troughs to larger scales
and causes more damping on smaller scales.
These features are helpful to
probe $\Delta \tau_{d}$,
as long as current and future CMB observational data
are accurate enough.
However, as an approximation,
this analytic result also has its limitation,
since the recombination history has been primarily
represented by only two parameters:
the recombination time $\tau_d$
and recombination width $\Delta\tau_d$ as an integrated effect.
Two different recombination histories
via different differential optical depth $q(\tau)$
would lead to the same amplitudes of bumps and troughs,
as long as they have same $\tau_d$  and $\Delta \tau_{d}$.}

\begin{figure}
 \resizebox{120mm}{!}{\includegraphics{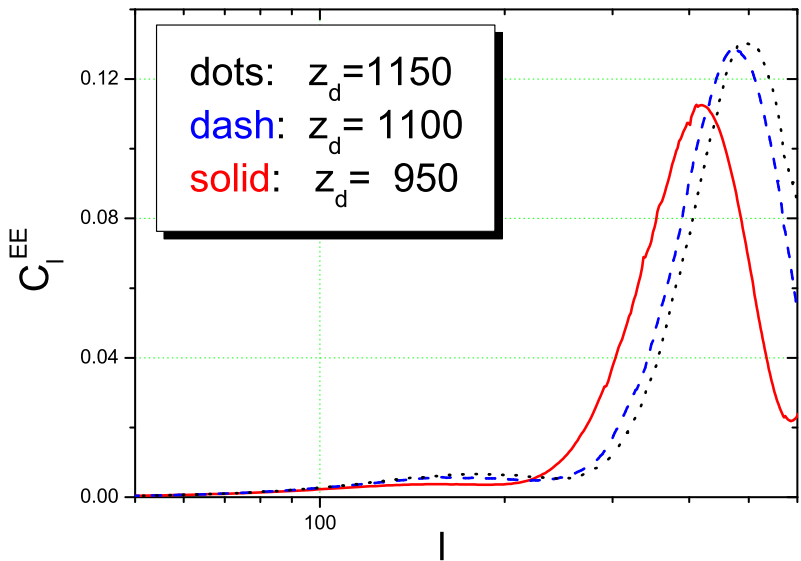}  }
\caption{  \label{fig14} \small
A late recombination time (larger $\tau_d$)
shifts the peaks and troughs of polarization
to larger angular scales.
 }
\end{figure}

Fig.\ref{fig14} shows that
a late recombination (greater $\tau_d$) shifts
the peaks and troughs of the polarization $C^{EE}_l$ to larger angular scales.
The property also holds for $C^{TT}_l$ and $C^{TE}_l$.
This can be explained  by the appearance of the function $j_l(k(\tau_0-\tau_d))$
as a factor in the analytic expressions of $a_l^T$ and $a_l^E$.

\begin{figure}
 \resizebox{120mm}{!}{\includegraphics{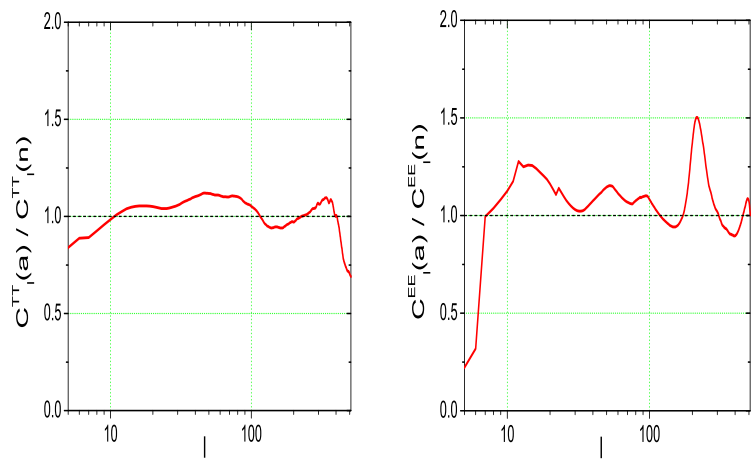} }
\caption{  \label{fig15} \small
The ratio of the analytic spectra to numerical spectra.
Left:  $C^{TT}_l(a)/ C^{TT}_l(n)$.
Right: $C^{EE}_l/(a) C^{EE}_l(n)$.
 }
\end{figure}

{ Fig.\ref{fig15} shows
the ratios of the analytic spectra to the numerical spectra,
$C^{TT}_l(a)/C^{TT}_l(n)$, and $C^{EE}_l(a)/C^{EE}_l(n)$.
The ratios are seen to be centered around $1$  for $l  \leq 500$,
showing a reasonable agreement between the analytic and numeric
on large angular scales. }

\begin{center}
{\em\Large 8. Conclusion and Discussions}
\end{center}

In this paper, we have presented an analytical calculation
of CMB anisotropies and polarization
generated by scalar metric perturbations in the synchronous gauge,
resulting in the explicit, analytic  expressions of the multipole moments
 $a^T_l$ in Eq.(\ref{aT2}) and $a^E_l$ in Eq.(\ref{aE3}),
and, thus, of all the analytical spectra $C^{TT}_l$, $C^{EE}_l$, and $C^{TE}_l$.
This has been implemented primarily through
an approximation treatment of time-integrations over the recombination process,
a technique used before for the case with RGW
as the generating source \cite{zhao,xia,Xia2}.
We have also dealt with
the removal of the residual gauge modes and
the joining condition  at the equality of radiation-matter
of the scalar perturbations.
Several approximations have been used,
such as the long wavelength approximation for scalar
perturbations during the RD era,
 tight-coupling approximation for the photons
during the recombination process.

These results are new and have significantly extended the
earlier preliminary works. The analytic expressions of
polarization $a^E_l$ and the related spectra $C^{EE}_l$, and
$C^{TE}_l$ are what have not been addressed in
Ref.\cite{Weinberg}.
Besides,  our analytic expression $a^T_l$ fulfils what
was not completed in Ref.\cite{Baskaran}, and,  to a great extent,
improves what was given in Ref.\cite{Weinberg},
as our expression
$a^T_l$ contains the separate contributions of monopole,
dipole, quadrupole, and Sachs-Wolfe terms.

Our analytic calculation shows that
the polarization $a^E_l$ is generated
mainly by the quadrupole of temperature anisotropies $\alpha_2$
via scattering.
Besides,
$a^E_l$ and of $\alpha_2$ are simultaneously generated
by the combination $ M(\tau_d) $,
so that
the resulting  $a^E_l$ and $\alpha_2$ have a similar structure
and both are smaller than the total temperature anisotropies $a^T_l$.

Furthermore,
the analytic expressions of $ a^E_l$  and $ a^T_l$ demonstrate explicitly
that
the peaks of the polarization $C^{EE}_l $
and of the temperature anisotropies $C^{TT}_l $ in $l-$ space
appear alternatingly.
These help to understand the important features of $C^{XX'}_l$.

As the major advantage of analytic expressions,
 $a_l^T$ and  $a_l^E$ explicitly show
the dependance upon the scalar perturbations,
 initial amplitude $C_0$,
  primordial spectrum index $n_s$,
   baryon fraction $\Omega_b$, damping factor $D(k)$, 
   recombination width $\Delta\tau_d$, and the recombination 
   time $\tau_d$. 
These properties are transparent in analytic expressions, but 
might not be directly obvious in the numerical code itself.
For instance,
the dependencies  upon $\Delta\tau_d$ tell that
a longer recombination process
yields a higher amplitude of polarization since $a_l^{E} \propto \Delta\tau_d$,
and brings  more damping of  $a_l^T$ and  $a_l^E$  on small scales
through $D_E(k) $, $ D_T(k)$.
The dependencies  upon $\tau_d$ tell that
a late recombination  shifts the peaks and troughs of spectra $C^{XX'}_l$
to larger angular scales.

The  spectra $C_l^{TT}$, $C_l^{TE}$ and $C_l^{EE}$
agree with the results
of the numerical codes
on large angular scales $l\lesssim 500 $,
covering the first two peaks and troughs of $C_l^{XX'}$.
On smaller scales,
the analytical spectra deviate considerably from the numerical ones,
as is expected for the long wavelength approximations.
Serving as
a complement to the numerical studies,
the preliminary analytical calculations
efficiently promote the analysis of effects
upon  $C^{XX'}_l$ by various physical processes,
and improve our understanding the important features of the observed CMB.

Based upon
the framework presented in this paper,
several points can be further improved for more accurate spectra $C_l^{XX'}$.
Some of them are listed as the following.
One can extend the analytical calculation to smaller scales \cite{Bartolo}.
For the solutions of perturbations $h$ and $\eta$,
Eqs.(\ref{h}) and (\ref{eta1}),
   one may include higher order terms in  $k\tau$.
Consistent with this,
one could do a finer treatment of the baryon component
before the recombination,
including the time dependence of the ratio $R(\tau)$ as in Eq.(\ref{dotR}).
One could also try to include the modifications from
the relativistic neutrino component
   during the RD era.
Finer examinations can be made on the initial condition during the RD era.
  For instance,  alternative forms could be tried for
  the slowly growing mode  $D$ other than that in Eq.(\ref{D}),
 and possible allowances could be tested for initial isocurvature perturbations
    besides the adiabatic ones.
Further examinations on the gauge modes for smaller scales
 could be made  during the MD era.
Very importantly, one should include the reionization
             occurred around a redshift $z\sim 11$,
    a process secondary only to the recombination.
This will definitely bring about modifications of  $C_l^{XX'}$
    on large angular scales $l\sim 5$ \cite{Xia2}.
Finally, to extract possible signals of RGW from observations,
  one should separate the contribution of RGW with various ratio $r$
  from scalar perturbations in the total spectra $C_l^{XX'}$,
which can be done within the framework in synchronous gauge
by using the results in this paper and our previous work on RGW.

 ~

\begin{center}
{\em\LARGE   Appendix:}
{\LARGE  The multipole moments for radiation field}
\end{center}

On a 2-dimensional unit sphere  with a metric
\begin{equation}
d\sigma ^2=g_{ab}dx^a dx^b=d\theta^2+\sin^2{\theta}d\phi^2,
\end{equation}
a general radiation field is usually characterized by the
following $2\times 2$
polarization tensor  \cite{chandrasekar,kamionkowski,Baskaran},
\be \label{Pab}
\begin{array}{l}
 P_{ab} = \frac{1}{2}\left( {\begin{array}{*{20}c}
  I+ Q                   &    -(U-iV)\sin{\theta}  \\
   -(U+iV)\sin{\theta}   &    (I-Q)\sin^2{\theta} \\
\end{array}} \right) \\
\end{array}
\ee with the four Stokes parameters ($I$, $Q$, $U$, $V$),
where $I$
is the intensity of radiation,
$Q$ and $U$ describe the linear polarization,
and $V$ is the circular polarization.
In the case of CMB,
the Thomson scattering during the recombination does not
generate the circular polarization \cite{chandrasekar},
so we set $V=0$.
Note that $I$ is a scalar on the 2-dim sphere
under the transformation of $\theta$ and $\phi$,
but $Q$ and $U$ transform among  themselves.
To deal with this problem,
several formulations have been proposed,
such as the total angular momentum method
using the spin-weighted spherical harmonic functions \cite{HuWhite},
and the spin raising and lowering operator method
\cite{zaldarriaga,kamionkowski}.
These two treatments are essentially equivalent,
and the latter  will be adopted in the following.
The tensor in Eq.(\ref{Pab}) consists of two parts:
\[
 P_{ab} = \frac{1}{2}Ig_{ab}+ P_{ab}^{STF},
\]
where $ \frac{1}{2}Ig_{ab}$ for
the temperature anisotropies
is of scalar nature,
and  $P_{ab}^{STF}$ for the polarization
is the symmetric trace-free (STF):
\be \label{STF}
\begin{array}{l}
 P_{ab}^{STF} = \frac{1}{2}\left( {\begin{array}{*{20}c}
   Q & -U\sin{\theta}  \\
   -U\sin{\theta} & -Q\sin^2{\theta} \\
\end{array}} \right), \\
\end{array}
\ee
from which one can construct two linear independent,
invariant quantities involving
its second order covariant derivatives \cite{Baskaran}:
\be  \label{E and B}
E(\theta, \phi)=-2(P_{ab}^{STF})^{;a;b}, \ \ \ B(\theta,
\phi)=-2(P_{ab}^{STF})^{;b;d}\epsilon^a_d,
\ee
where
\be
\begin{array}{l}
\epsilon_{ab} = \left( {\begin{array}{*{20}c}
   0 & -\sin{\theta}  \\
   \sin{\theta} & 0 \\
\end{array}} \right) \\
\end{array}
\ee
is a completely antisymmetric pseudo-tensor.
$E$ is a scalar on the 2-sphere
and  $B$ is a pseudo-scalar.
{ It is revealing to write $P_a \equiv (P_{ab}^{STF})^{;b}$.
Then $E=P_a\, ^{;a}$ is a divergence of $P_a$,
 and $B=P_{a;b}\epsilon^{ab}$ is a curl of $P_a$.
In this regard,  $E$ is referred to as the ``electric'' polarization,
and $B$ as the ``magnetic'' polarization.}
As can be checked,
by directly calculating the covariant derivatives
on the 2-sphere, one has   \cite{kamionkowski,zaldarriaga}
\be  \label{E}
E=-\frac{1}{2}[\baredth^2(Q+iU)+ \edth^2(Q-iU)],
\ee
\be \label{B}
B=\frac{i}{2}[\baredth^2(Q+iU)- \edth^2(Q-iU)],
\ee
where $\edth ^2$ is the raising operator
acting twice,
and  $\baredth ^2$
is the lowering operator acting twice,
\be
\edth ^2 \    (Q-iU)(\mu ,\phi ) = ( -
        \partial _\mu   - \frac{-i\partial_{\phi}}{{1 - \mu ^2 }})^2
                 [(1 - \mu ^2 )  (Q-iU)(\mu ,\phi )],
\ee
\be
\baredth ^2 \  (Q+iU)(\mu ,\phi )
 = ( - \partial _\mu + \frac{-i\partial_{\phi}}{{1 - \mu ^2 }})^2
                 [(1 - \mu ^2 )(Q+iU)(\mu ,\phi )],
\ee
where $\mu=\cos \theta$.

Since  $I$, $E$, and $B$ are scalars on the 2-sphere,
they can be expanded in terms of the spherical harmonics
$Y_{lm}(\theta, \phi)$ as a complete and orthonormal basis \cite{Baskaran}:
\be \label{Texpand}
I(\theta, \phi)=
\sum\limits_{l=0}\limits^{\infty}
\sum\limits_{m=-l}\limits^{l}
 a_{lm}^T Y_{lm}(\theta, \phi),
\ee
\be \label{Eexpand}
E(\theta, \phi)=
\sum\limits_{l=2}\limits^{\infty}
\sum\limits_{m=-l}\limits^{l}
\left[\frac{(l+2)!}{(l-2)!} \right]^{\frac{1}{2}}a_{lm}^E Y_{lm}(\theta, \phi),
\ee
\be \label{Bexpand}
B(\theta, \phi)=\sum\limits_{l=2}\limits^{\infty}
\sum\limits_{m=-l}\limits^{l}
\left[\frac{(l+2)!}{(l-2)!}
   \right]^{\frac{1}{2}}a_{lm}^B Y_{lm}(\theta, \phi).
\ee
For technical simplicity,
one can choose the  coordinate with the polar
axis $\bf \hat z$ pointing along the wave vector $\bf k$
of the scalar perturbation mode:  $\bf k || \hat z$.
{
Let an unpolarized incident light have an intensity $I'$
scattered on a charge.
Using the differential Thomson scattering cross-section
$\frac{d\sigma}{d\Omega}=\frac{3\sigma_T}{8\pi}|\epsilon'\cdot \epsilon|$,
with $\epsilon'$ and $ \epsilon$ being
the polarization of incident and outgoing light, respectively,
one obtains   \cite{Kosowsky}
$I=\frac{3\sigma_T}{16\pi}(\theta)I'(1+\cos^2\theta)$,
 $Q=\frac{3\sigma_T}{16\pi}(\theta)I'\sin ^2\theta $, and
 $U=0$
for the outgoing wave,
where  $\theta$ is
the angle between the incident and outgoing directions.
The result does not depend on the azimuthal angle $\phi$.
As a corollary, in an azimuthal symmetric configuration,
Thomson scattering of an unpolarized light yields
\be
I=I(\theta), \,\,\,\,\,  Q=Q(\theta), \,\,\,\,\, U=0.
\ee
for the outgoing wave.
This is just the situation with a $\bf k$-mode of
density perturbations at the last scattering.
As explained in Section 2,
The $\bf k$-mode  of density perturbation is
azimuthal symmetric about the $\bf k$ axis.
At the last scattering,
the incident light is unpolarized.
Therefore, in Eq.({\ref{n1}})
we only need $I$ and $Q$ for
a $\bf k$ mode of the density perturbations
\cite{Kosowsky,zaldarriaga,kamionkowski}.
}

Since $Q$ only depends on $\theta$,
so that $\baredth ^2 Q = \edth ^2 Q =
\partial _\mu ^2 [(1 - \mu ^2) Q(\mu )]$,
resulting
\be \label{QE}
E=-\partial _\mu ^2 [(1 - \mu ^2) Q(\mu )],
\ee
\be
B=0,
\ee
i.e., the scalar metric perturbations generate no
polarization of magnetic type  \cite{zaldarriaga}.
{
Another more geometric way to see why $B=0$ and $E\ne 0$ is to
use the definitions in Eq.(\ref{E and B}).
Since $U=0$ for a $\bf k$-mode of density perturbation,
the polarization matrix in Eq.(\ref{STF}) reduces to
\be
\begin{array}{l}
 P_{ab}^{STF} = \frac{1}{2}\left( {\begin{array}{*{20}c}
   Q & 0  \\
   0 & -Q\sin^2{\theta} \\
\end{array}} \right), \\
\end{array}
\ee
and, a direct calculation yields
\be
E=-Q_{,\theta\theta}-\frac{\cos\theta}{\sin\theta} Q_{,\theta},
\ee
\be
B=
\frac{2}{\sin\theta} Q_{,\theta\phi}
+\frac{2\cos\theta}{\sin^2\theta}Q_{,\phi}.
\ee
This tell us that the magnetic type of polarization $B$
essentially involves the derivative of $Q$ with respect to $\phi$,
and is a measure of asymmetry of polarization field
under the rotation about the $\bf k$ axis.
Since $Q$ is independent of $\phi$, one has $B=0$.}

{ It is interesting to compare with the case GW,
where the rotational symmetry 
is lost for the $\bf k$ mode of GW,
and the outgoing light after Thomson scattering would
be a general linear polarized one,
with all three Stokes parameters
$I=I(\theta,\phi)$, $ Q=Q(\theta,\phi)$, and $  U=U(\theta,\phi)\neq 0$,
depending on  $\theta  $ as well as  $\phi$ \cite{basko,Polnarev,zhao},
and resulting in $E\ne 0$ and $B\ne 0$.
This distinguished feature of non-vanishing magnetic type of polarization
of CMB can be served as a possible channel to detect
gravitational waves. }

The multipole moments  $a_{lm}^T$
of temperature anisotropies
and  $a_{lm}^E$ of the electric type of polarization are given by
\be \label{alm}
a_{lm}^T(k)
   =2\pi \int_{-1}^{1} {d\mu }\,  Y^*_{lm}(\mu)  I (\tau ,\mu ),
\ee
\be
a_{lm}^E(k)
   =2\pi
   \left[\frac{(l-2)!}{(l+2)!} \right]^{\frac{1}{2}}
   \int_{-1}^{1}d\mu \,    Y^*_{lm}(\mu) E(\mu).
\ee
Both $a_l^{T}$ and $a_l^E$ are observables
on the sky.
Since $I $ and $E$ are now functions of $\theta$ only,
one can set the magnetic index $m=0$ in the above expressions
and uses the replacements
 $Y_{l0}(\theta) =  \sqrt{\frac{2l+1}{4\pi}}  P_l(\mu)$
 and $a^T_{lm}\rightarrow a^T_{l0}$ and  $a^E_{lm}\rightarrow a^E_{l0}$.

Firstly, we calculate  the multipole moments $a^T_{l0}$
at the present time $\tau_0$.
From Eq.(\ref{alm}),
using Eq.(\ref{Ialpha}) and Eq.(\ref{alpha_n}),
one has
\ba  \label{atlm}
&&a_{l0}^T(k)  = 2\pi \gamma \sqrt{\frac{2l+1}{4\pi}}
     \int_{-1}^{1} {d\mu }  P_l(\mu)
         \,  \alpha_k(\tau_0 ,\mu ) \nonumber\\
&=& 2\pi \gamma  \sqrt{\frac{2l+1}{4\pi}} \int_{-1}^{1} {d\mu } P_l(\mu)
   \int_0^{\tau_0}  {d\tau}
      e^{ - \kappa(\tau_0 ,\tau ) - i  \mu k(\tau_0 -\tau )}
           \left[ \frac{dH}{d\tau }  - \mu ^2 \frac{{dH_l }}{{d\tau }}
     + q(\mathcal{I} _1
     + i\mu v_b
     - \frac{1}{2}P_2 (\mu )\mathcal{I}_2 )\right]. \nonumber
\ea
Making use of the relation
\be \label{bessel}
\int_{-1}^1 {d\mu } P_l(\mu)e^{ - i\mu x}
   =2  ( - i)^l j_l (x),
\ee
the above expression of $a_{l0}^T$ is reduced to
\ba \label{alT}
&&a^T_{l0}(k)= \gamma(-i)^l \sqrt {4\pi (2l + 1)}a^T_l(k),
\ea
where
\be \label{alt}
a_l^T(k)=\int_0^{\tau_0} {d\tau  }
\left[ e^{ - \kappa(\tau)}
 (\frac{{dH}}{{d\tau }} + \frac{{dH_l }}{{d\tau
}}\frac{{d^2 }}{{d\zeta ^2 }}) + V(\tau )(\mathcal{I} _1 - v_b
\frac{d}{{d\zeta }} - \frac{{3}}{4}\mathcal{I} _2(1+\frac{{d^2
}}{{d\zeta ^2 }}) )\right]j_l(\zeta),
\ee
with the variable $ \zeta \equiv k(\tau_0  - \tau )$.

Next, we calculate  the multipole moments $a^E_{l0}$
at the present time $\tau_0$.
\be \label{2}
a_{l0}^E(k)
   =2\pi \left[\frac{(l-2)!}{(l+2)!} \right]^{\frac{1}{2}}
       \sqrt{\frac{2l+1}{4\pi}}    \int_{-1}^{1}d\mu  P_{l}(\mu) E(\mu).
\ee
By Eq.(\ref{Qbeta}) and Eq.(\ref{QE}), one has
\be \label{Ebeta}
E(\mu)
     =-\gamma\partial _\mu ^2 [(1 - \mu ^2) \beta_k(\mu )].
\ee
Substituting the expression $\beta_k$ of Eq.(\ref{beta_n})
into Eq.(\ref{Ebeta}) yields
\ba \label{En}
E (\mu ) &= &  - \gamma\frac{3}{4}\int_0^{\tau_0}  {d\tau  V(\tau )}
     \mathcal{I} _2 (\tau )\partial _\mu  ^2 [(1 - \mu ^2 )^2
    e^{ -     i\zeta \mu } ] \nonumber\\
     &=&  \gamma \frac{3}{4}\int_0^{\tau_0} {d\tau  V(\tau )}
        \mathcal{I}_2 (\tau)(1 + \partial _\zeta  ^2 )^2
                 (\zeta ^2 e^{ - i\zeta \mu } ).
\ea
Substituting Eq.(\ref{En}) into Eq.(\ref{2})
and using Eq.(\ref{bessel}), one has
\ba \label{3}
a^E_{l0}(k) & = & \gamma 2\pi \sqrt{\frac{2l+1}{4\pi}}\frac{3}{4}
   \left[\frac{(l - 2)!}{(l +2)!}\right]^{\frac{1}{2}}
      \int_0^{\tau_0}  d\tau \int^1_{-1} d\mu
   P_{l}( \mu )
   V(\tau )\mathcal{I}_2 (\tau )(1 +
   \partial _\zeta  ^2 )^2 (\zeta^2 e^{ - i\zeta \mu } ) \nonumber\\
&=&  \gamma(-i)^l\sqrt{4\pi(2l+1)} \frac{3}{4}
   \left[\frac{{(l + 2)!}}{{(l- 2)!}} \right]^{\frac{1}{2}}
   \int_0^{\tau_0} d\tau
   V(\tau ) \mathcal{I}_2 (\tau )
     (1 +\partial _\zeta  ^2 )^2 (\zeta ^2 j_l (\zeta )).
\ea
Using the relation for the  spherical-Bessel functions
\[
j_l^{\,''} (x) + 2\frac{{j_l ^{\,'}(x)}}{x}
         + [1 - \frac{{l(l + 1)}}{{x^2 }}]j_l (x) = 0
\]
to replace $ j_l^{\,''}(\zeta ) $,
the term $(1 + \partial _\zeta ^2 )^2(\zeta ^2 j_l (\zeta )) $
in Eq.(\ref{3}),
one obtains
\be \label{alE}
a^E_{l0}(k) =  \gamma(-i)^l\sqrt{4\pi(2l+1)}a_l^E(k),
\ee
where
\be \label{ale}
a_l^E(k)=\frac{3}{4}\left[\frac{{(l + 2)!}}{{(l -2)!}} \right]^{\frac{1}{2}}
   \int_0^{\tau_0} d\tau
      V(\tau) \mathcal{I}_2 (\tau )\frac{{j_l (\zeta )}}{{\zeta ^2 }}.
\ee
One can check that  $a_l^T$ and $a_l^E$
in Eqs.(\ref{alt}) and (\ref{ale})
are essentially $\alpha_k$ and $\beta_k$
projected on the basis $P_l(\mu)$,
respectively:
\ba   \label{aT multipole}
  && a_l^T(k)=  i^{l}\frac{1}{2}\int_{-1}^{1}{d\mu }
     P_l \cdot \alpha _k (\tau ,\mu ),\\
 && a_l^E(k)=  i^{l}\frac{1}{2}\int_{-1}^{1}{d\mu }
     P_l \cdot \beta_k (\tau ,\mu ).  \label{aE multipole}
\ea
The main result of the Appendix
is the expressions of $a^T_l$ in Eq.(\ref{alt}) and
$a^E_l$ in Eq.(\ref{ale}),
which have been used in Section 5.

\

ACKNOWLEDGMENT:
Z. Cai has been partially supported by
National Science Fund for Fostering Talents in Basic Science (J0630319), 
and Z. Cai would like to thank Prof. Li-Zhi Fang for his encouragements and
 thank Dr. Xiaohui Fan for partial supports. 
Y. Zhang's research work is supported by the CNSF No.11073018,
               SRFDP, and CAS.

%\newpage

\newpage

\end{document}